\documentclass[sigplan,screen]{acmart}
\setcopyright{none}
\renewcommand\footnotetextcopyrightpermission[1]{}
\usepackage{multirow}

\usepackage{graphicx}
\usepackage[caption=false]{subfig} 
\usepackage{hyperref}

\makeatletter
\renewcommand{\p@subfigure}{\figurename~\thefigure}
\makeatother

\usepackage{enumitem}

\AtBeginDocument{%
  }

\acmConference[]{}{}

\acmSubmissionID{3144}

\begin{document}

\title{Low-Barrier Dataset Collection with Real Human Body for Interactive Per-Garment Virtual Try-On}

\author{Zaiqiang Wu}
\affiliation{%
 \institution{The University of Tokyo}
 \country{Japan}
}

\author{Yechen Li}
\affiliation{%
 \institution{The University of Tokyo}
 \country{Japan}
}

\author{Jingyuan Liu}
\affiliation{%
 \institution{The University of Tokyo}
 \country{Japan}
}

\author{Yuki Shibata}
\affiliation{%
 \institution{SoftBank Corp}
 \country{Japan}
}
\author{Takayuki Hori}
\affiliation{%
 \institution{SoftBank Corp}
 \country{Japan}
}
\author{I-Chao Shen}
\affiliation{%
 \institution{The University of Tokyo}
 \country{Japan}
}
\author{Takeo Igarashi}
\affiliation{%
 \institution{The University of Tokyo}
 \country{Japan}
}

\renewcommand{\shortauthors}{Wu et al.}

\begin{abstract}
Existing image-based virtual try-on methods are often limited to the front view and lack real-time performance. While per-garment virtual try-on methods have tackled these issues by capturing per-garment datasets and training per-garment neural networks, they still encounter practical limitations: (1) the robotic mannequin used to capture per-garment datasets is prohibitively expensive for widespread adoption and fails to accurately replicate natural human body deformation; (2) the synthesized garments often misalign with the human body. To address these challenges, we propose a low-barrier approach for collecting per-garment datasets using real human bodies, eliminating the necessity for a customized robotic mannequin. We also introduce a hybrid person representation that enhances the existing intermediate representation with a simplified DensePose map. This ensures accurate alignment of synthesized garment images with the human body and enables human-garment interaction without the need for customized wearable devices. We performed qualitative and quantitative evaluations against other state-of-the-art image-based virtual try-on methods and conducted ablation studies to demonstrate the superiority of our method regarding image quality and temporal consistency. Finally, our user study results indicated that most participants found our virtual try-on system helpful for making garment purchasing decisions.

\end{abstract}


\begin{teaserfigure}
  \includegraphics[width=\textwidth]{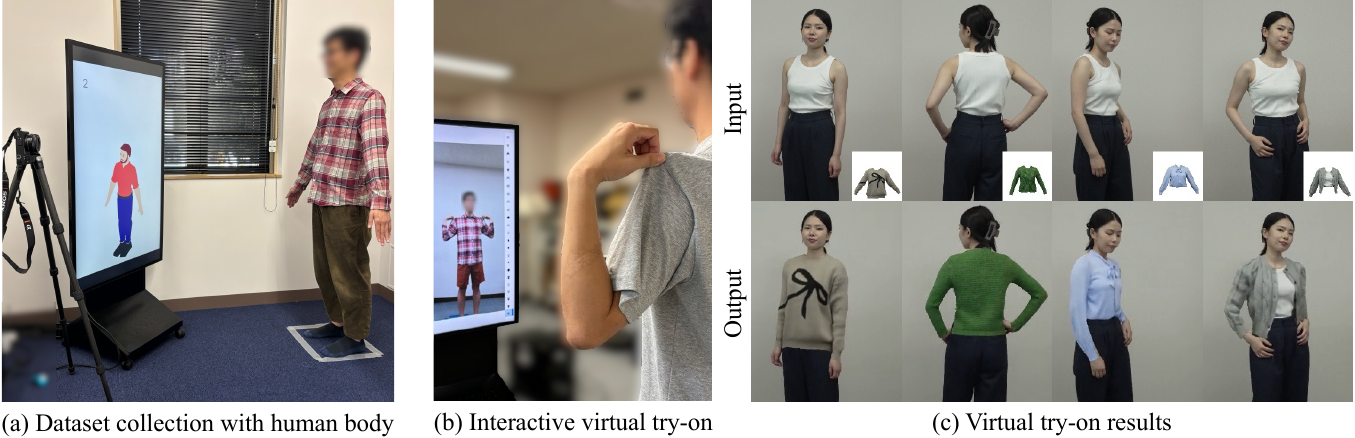}
  \caption{
  (a) Our low-barrier dataset collection method requires only a camera, a monitor, and a human body wearing the target garment. No customized devices are required. (b) Our system is efficient, achieving a frame rate of approximately $8$ frames per second on a moderate PC, thereby supporting interactive virtual try-on experiences. (c) Our system can generate highly plausible virtual try-on results for various target garments from different viewpoints, including side and back.}
  \Description{}
  \label{fig_teaser}
\end{teaserfigure}



\maketitle

\section{Introduction}
\label{sec:intro}
Virtual try-on technology, which enables users to try on garments virtually without needing physical access to them, has recently garnered significant attention from researchers. Existing image-based virtual try-on methods~\cite{jetchev2017conditional,dong2019fw,choi2021viton,lee2022high, fang2024vivid, karras2024fashionvdmvideodiffusionmodel} use a 2D image of a target garment to create realistic virtual try-on results. However, these methods face several limitations that hinder their practical application in real-world scenarios. Firstly, they often produce poor results for general users because they are typically trained on biased datasets that include only fashion models and a limited range of poses. Secondly, their complexity prevents them from running in real-time, significantly impacting the virtual try-on experience.

\begin{figure}[t!]
    \centering\includegraphics[width=\linewidth]{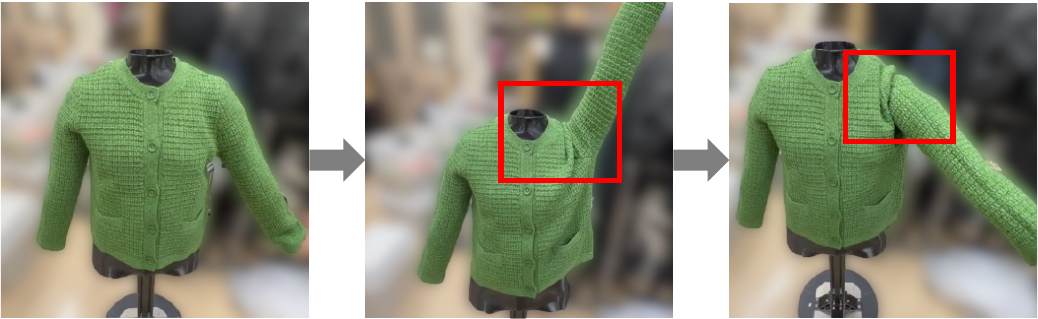}
    \caption{Unnatural wrinkles of the target garment occur around the shoulder region when collecting per-garment datasets using a customized robotic mannequin.}
    \label{fig_robotic}
\end{figure}

A subfield of image-based methods, per-garment virtual try-on methods~\cite{chong2021per, wu2024virtual}, addresses these challenges by collecting garment-specific datasets and training garment-specific neural networks. These methods leverage per-garment datasets collected using a robotic mannequin, enabling the capture of massive paired images of the target and measurement garments in identical poses.
Despite the advantages of using a robotic mannequin, several limitations persist: 
\begin{itemize}
	\item The robotic mannequin fails to replicate the natural human body dynamics and the garment-skin interactions. This results in unnatural wrinkles appearing on the target garment, as shown in \autoref{fig_robotic}.
	\item The high cost of the robotic mannequin limits its widespread use and poses a major barrier to dataset collection.
	\item Significant differences in appearance between the robotic mannequin and real humans hinder the application of existing human-centric vision models for generating annotations to guide alignment, resulting in misalignment between the synthesized garment and the human body.
\end{itemize}

To address these limitations, we propose a low-barrier dataset collection method that uses real human bodies instead of a customized robotic mannequin. Our approach requires only a camera, a monitor, and a human model wearing the target garment, as shown in~\autoref{fig_teaser}(a). Non-experts can create their own virtual try-on model for a specific garment by physically wearing it and demonstrating how it looks under various poses to a camera, which can be regarded as a form of human-in-the-loop machine learning.
Our experiments indicate that it takes approximately two minutes for a real human, without prior professional training, to complete the task, whereas using a customized robotic mannequin requires around two hours~\cite{chong2021per}.

Additionally, using real humans allows us to enhance the prior intermediate representation~\cite{wu2024virtual} with a simplified DensePose~\cite{guler2018densepose} map, ensuring precise alignment between the synthesized garment and the human body. Furthermore, DensePose's ability to capture rough garment deformation allows users to interact with the synthesized garment by pulling on the garment they are wearing, eliminating the need for a customized measurement garment as required by~\cite{chong2021per}. Moreover, the flexibility of real humans facilitates the collection of datasets for a wider variety of garment types and poses, enabling our methods to generate plausible results for various garments across diverse poses, as illustrated in \autoref{fig_teaser}(c). 

Our method achieves a frame rate of approximately 8 frames per second on a moderate PC, enabling an interactive virtual try-on experience, as shown in \autoref{fig_teaser}(b). Both qualitative and quantitative evaluations indicate that our method surpasses existing approaches in terms of visual quality and temporal consistency.

Our contributions can be summarized as follows: 
\begin{itemize} 
	\item We propose a low-barrier dataset collection approach that eliminates the need for expensive customized robotic mannequins, significantly reducing the cost of per-garment dataset collection. 
	\item We enhance the virtual measurement garment~\cite{wu2024virtual} by incorporating a simplified DensePose map, ensuring precise human-garment alignment and facilitating human-garment interaction without requiring customized wearable devices. 
	\item We perform extensive qualitative and quantitative evaluations, along with a user study, to assess the usability and utility of our virtual try-on system.       
\end{itemize}
\section{Related Work}
\label{sec:related}
\subsection{3D model-based Virtual Try-on}
3D model-based virtual try-on methods~\cite{guan2012drape,sekine2014virtual,santesteban2019learning,patel2020tailornet,pan2022predicting} involve creating a digital representation of the human body and developing 3D models of clothing items. 
The try-on results are generated by simulating the draping, fitting, and movement of the clothing items on the 3D body model, utilizing either physically-based simulations~\cite{cirio2014yarn,kaldor2008simulating,narain2012adaptive,selle2008robust} or data-driven methods~\cite{casado2022pergamo,grigorev2023hood,halimi2023physgraph,lahner2018deepwrinkles,pan2022predicting,patel2020tailornet,santesteban2019learning,santesteban2022snug,santesteban2021self,xiang2021modeling}.
Despite their ability to realistically capture garment deformations across diverse poses and viewpoints, these methods face two critical limitations: (1) They require time-consuming manual 3D modeling for each garment, which significantly hinders scalability; (2) They primarily generate try-on results for digital avatars rather than real users, leading to substantial misalignment when overlaying the 3D garment on human bodies, thereby reducing their effectiveness for personalized purchase decisions.
In contrast, our method can automatically generate the necessary data for each garment item with minimal human intervention and ensure precise alignment between the synthesized garment and the human body without the need for depth sensors or wearable devices.

\subsection{Image-based Virtual Try-on}
Image-based virtual try-on methods aim to synthesize a realistic image of a person wearing a specific garment using a combination of the person's image and an in-shop garment image~\cite{song2023image}. 
Unlike 3D model-based methods, 2D image-based methods do not need manual 3D modeling, making them more accessible while still achieving a realistic appearance of the target garment. 
Some early works~\cite{jetchev2017conditional,choi2021viton,lee2022high} are based on Generative Adversarial Networks (GAN)~\cite{goodfellow2014generative} and its variations~\cite{zhu2017unpaired,karras2019style,karras2020analyzing}. 
More recently, many recent works~\cite{kim2024stableviton,wang2024mv,xu2024ootdiffusion} have leveraged diffusion models to improve image quality in virtual try-on methods. 

While these methods generate realistic static try-on images, users often prefer to see try-on videos that showcase a person dressing and moving naturally.
Initial approaches, such as FW-GAN~\cite{dong2019fw}, FashionMirror~\cite{chen2021fashionmirror}, and ClothFormer~\cite{jiang2022clothformer}, incorporate optical flow for garment image warping to maintain temporal consistency over time. 
More recent methods, including ViViD~\cite{fang2024vivid}, Tunnel Try-on~\cite{xu2024tunnel}, and Fashion-VDM~\cite{karras2024fashionvdmvideodiffusionmodel}, utilize image and video diffusion models to enhance try-on quality. 

Typically, these methods train a universal network to accommodate all types of garments using human identity-aware representations and target garment images.
As a result, they need a large dataset that includes a diverse range of human identities and garment styles to ensure generalizability.
Moreover, these methods face challenges such as slow processing speeds, data bias towards front-view fashion models, and issues with altering human identity.
In contrast, our method focuses on generating high-fidelity and temporal consistent garments from multiple viewpoints while preserving the user's human identity.

\subsection{Per-garment Virtual Try-on} 
Per-garment virtual try-on methods achieve superior try-on results for specific target garments by training garment-specific networks. The pioneer work by~\citet{chong2021per} employed a robotic mannequin to collect per-garment datasets and trained image-to-image translation networks. These networks map the images of a physical measurement garment to the images of the target garment. While this method has demonstrated superior try-on results for specific garments, it requires users to wear a customized measurement garment, limiting its application.
\citet{wu2024virtual} proposed an alternative method that estimates a 3D human pose and deforms a virtual measurement garment mesh accordingly, serving as an intermediate representation and eliminating the need for a physical measurement garment. However, this approach suffers from noticeable misalignment between the synthesized garment and the human body.
Most importantly, both approaches rely on using a customized robotic mannequin for dataset collection, which hinders their reproducibility.
Our method overcomes this limitation by capturing per-garment datasets using real human bodies, significantly reducing the barrier to data collection.

\subsection{Human-in-the-loop Machine Learning}
Human-in-the-loop machine learning aims to enhance the performance of machine learning models by incorporating human expertise~\cite{wu2022survey}. Interactive Machine Teaching~\cite{ramos2020interactive,simard2017machine}, a subfield of human-in-the-loop machine learning, aims to help non-experts easily create their own machine learning model without the need for technical skills. It aims to bridge the gap between domain knowledge holders and machine learning technology, making the model creation process more accessible. For instance, Teachable Machine~\cite{carney2020teachable} allows novice users, without any machine learning skills, to transfer their knowledge of object classification to a machine learning model by presenting various views of each object to a camera. To reduce the burden on users and make it easier to highlight the object of interest, LookHere~\cite{zhou2022gesture} employs users' deictic gestures to annotate the region of interest.

Our method can be considered as a variant of human-in-the-loop machine learning. It enables novice users, without machine learning skills or specialized devices, to create a virtual try-on model by physically wearing the target garment and demonstrating how it looks under various poses to a camera. Unlike typical systems that rely on human knowledge for annotation or instruction for training, our system leverages the users' physical bodies.

\begin{figure}[h!]
    \centering
    \subfloat[Training/inference pipeline of~\citet{wu2024virtual}.]{%
        \includegraphics[width=\linewidth]{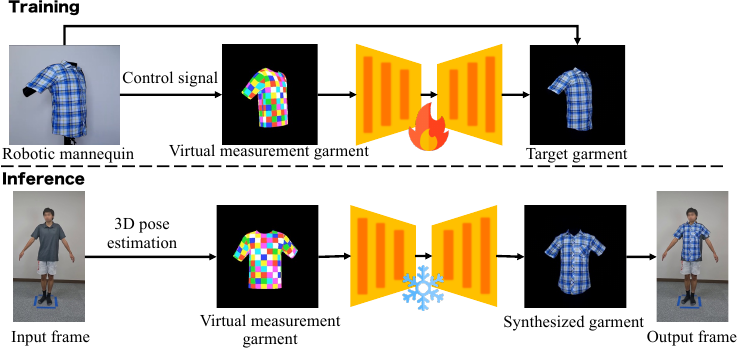}
        \label{fig_virtual_overview}
    }
    \vfill
    \subfloat[Training/inference pipeline of our method.]{%
        \includegraphics[width=\linewidth]{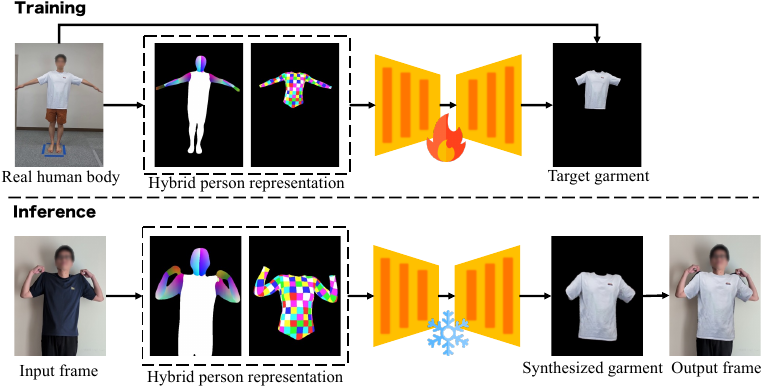}
        \label{fig_ours_overview}
    }
    \caption{
    Training/inference pipeline comparison: (a) prior per-garment method~\cite{wu2024virtual} vs. (b) our method.
    }
    \label{fig_comparison_pipelines}
\end{figure}


\begin{figure}[h!]
    \centering
    \includegraphics[width=\linewidth]{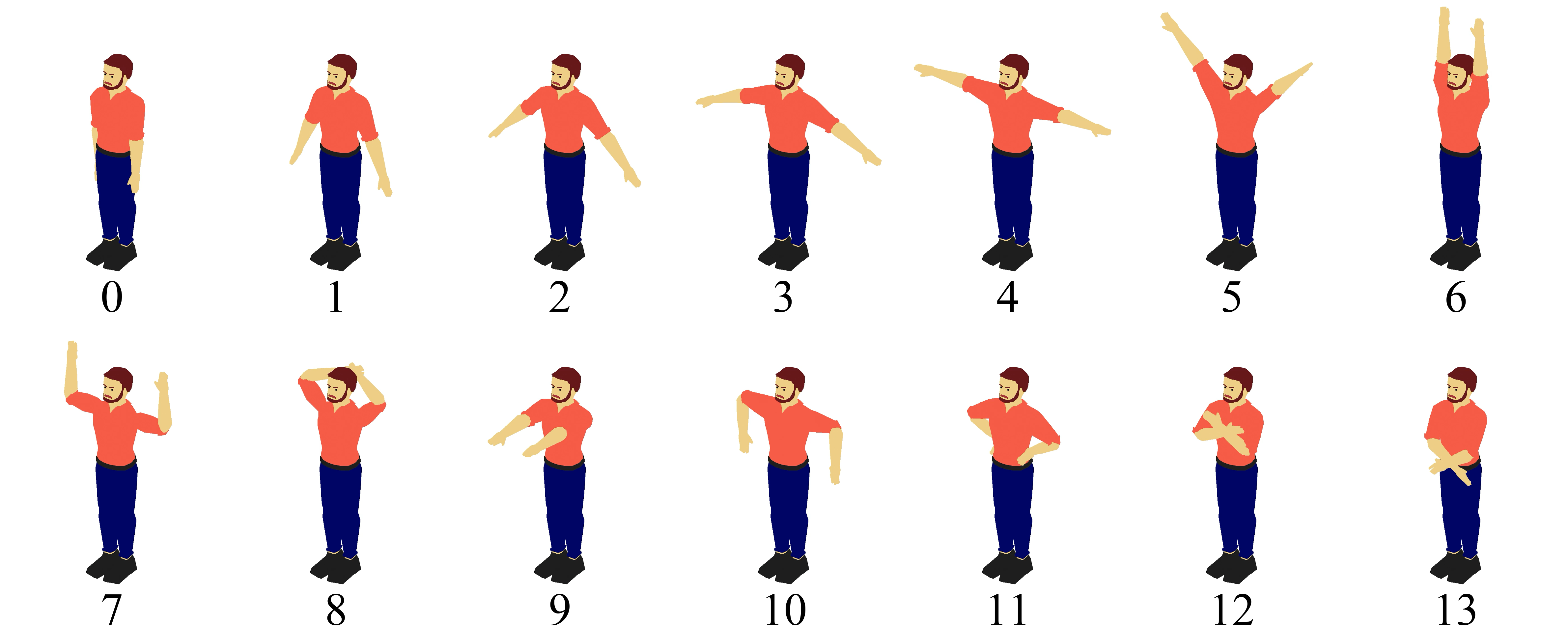}
    \caption{The predefined pose set we used during the data capturing procedure.}
    \label{fig_allposes}
\end{figure}

\begin{figure}[h!]
    \centering
    \includegraphics[width=\linewidth]{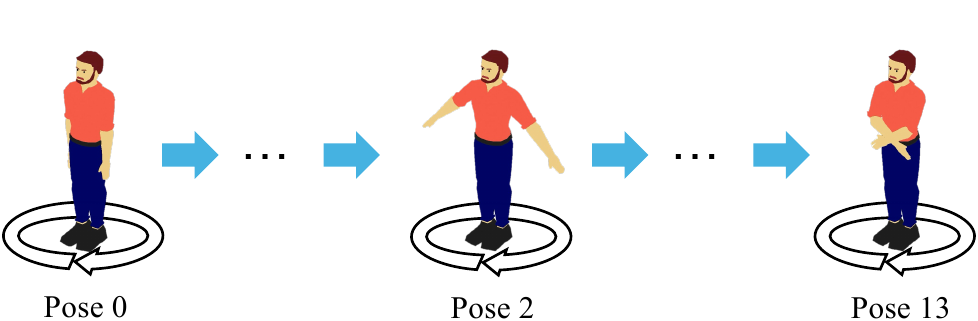}
    \caption{
    The data capturing procedure: The person replicates each predefined pose, rotating 360 degrees slowly in place for each pose.
    }
    \label{fig_movements}
\end{figure}

\begin{figure}[hbt!]
    \centering
    \includegraphics[width=\linewidth]{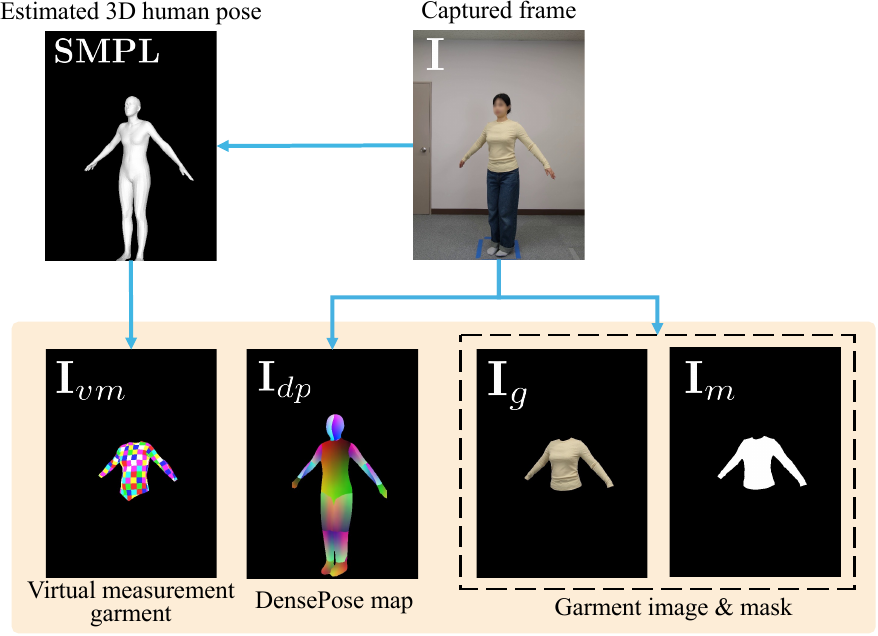}
    \caption{
    Per-garment dataset generation from recorded video of a person performing predefined poses.}
        \label{fig_dataset}
\end{figure}

\begin{figure*}[h!]
    \centering
    \includegraphics[width=\linewidth]{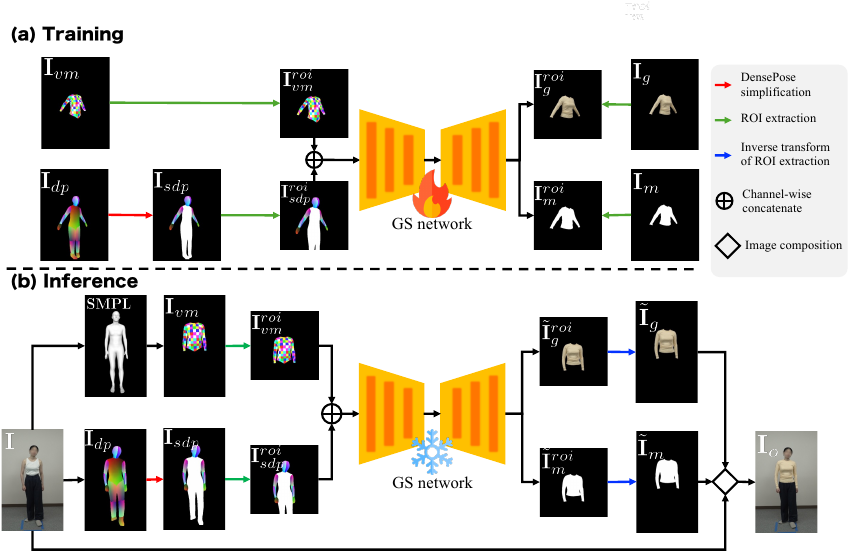}
    \caption{
    The details of the (a) training and (b) inference stages of our proposed per-garment virtual try-on method.}
    \label{fig_training_inference}
\end{figure*}

\begin{figure}[h!]
    \centering
    \subfloat[\citet{chong2021per}]{%
        \includegraphics[height=0.5\linewidth]{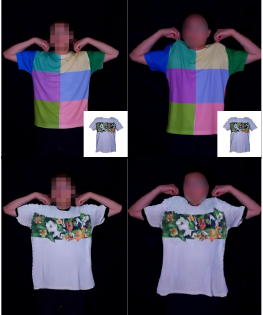}
    }
    \hspace{-0.1cm}
    \subfloat[Ours]{%
        \includegraphics[height=0.5\linewidth]{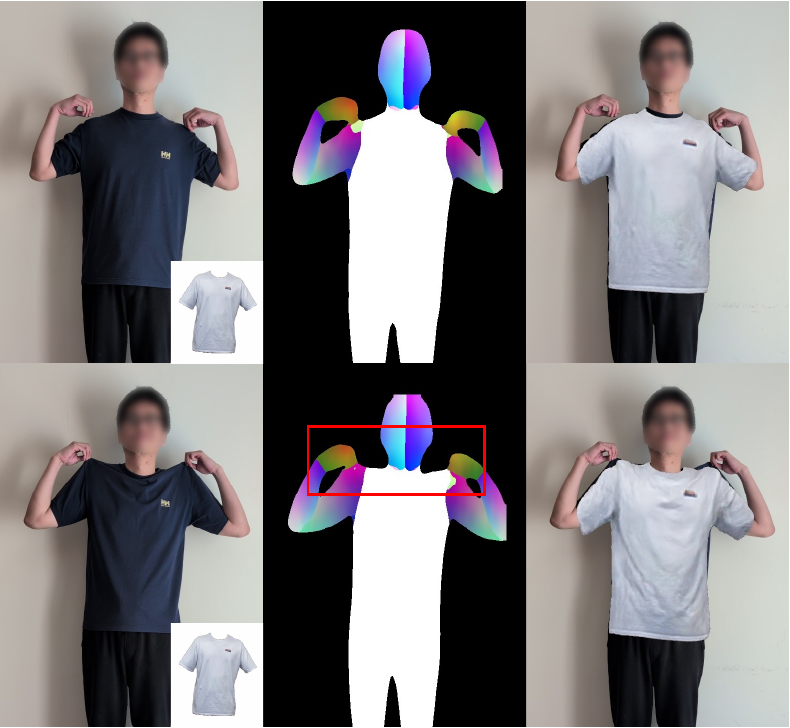}
    }
    \caption{
    Human-garment interaction: (a) prior per-garment method (\citet{chong2021per}) vs. (b) our method.
    \citet{chong2021per} requires a customized measurement garment, whereas our method does not require such a garment, thanks to the incorporation of the simplified DensePose map.
    }
    \label{fig_interaction}
\end{figure}

\section{Method}
\label{sec:method}

Our method builds upon the previous per-garment virtual try-on approach by \citet{wu2024virtual} and follows a similar image-to-image translation workflow. As illustrated in~\autoref{fig_comparison_pipelines}(a), during training, \citet{wu2024virtual} utilized a robotic mannequin and its control signals to generate paired images of the virtual measurement garment, which only contained pose information, and the target garment for training the per-garment network. During inference, they extracted the virtual measurement garment from the input frame using 3D pose estimation and fed it to the network to generate the try-on result. However, the absence of alignment information led to misalignment issues in their results (\autoref{fig_alignment_comparison}(a)).

In contrast, our method employs real human bodies to generate per-garment datasets for training the per-garment networks, as depicted in~\autoref{fig_comparison_pipelines}(b). We propose to use hybrid person representation $\mathbf{I}_{hybrid}\in \mathbb{R}^{6\times H\times W}$ as the intermediate representation:

\begin{align}\mathbf{I}_{hybrid}=\mathbf{I}_{vm}\oplus \mathbf{I}_{sdp}
\end{align}
where $\oplus$ denotes channel-wise concatenate. In this representation, $\mathbf{I}_{vm}$ denotes the image of the virtual measurement garment, which provides the annotation for the 3D human pose, while $\mathbf{I}_{sdp}$ denotes the simplified DensePose~\cite{guler2018densepose} map, which provides the annotation for 2D image space alignment.

To obtain $\mathbf{I}_{vm}$ from a person image $\mathbf{I}$, we first estimate its 3D pose represented by a SMPL~\cite{loper2015smpl} mesh. 
We then remove the unnecessary body parts (head, hands, and lower body) and texture the remaining parts with a grid pattern, and render it into a virtual measurement garment image $\mathbf{I}_{vm}\in \mathbb{R}^{3\times H\times W}$.
Unlike~\cite{wu2024virtual}, our method employs the SMPL mesh with full arms, which allows our method to handle both short-sleeved and long-sleeved garments.

To obtain $\mathbf{I}_{sdp}$ from a person image $\mathbf{I}$, we estimate the DensePose map $\mathbf{I}_{dp}$ from the input image. We then simplify the DensePose map by painting the torso and lower body parts white, resulting in the simplified DensePose map $\mathbf{I}_{sdp} \in \mathbb{R}^{3\times H\times W}$: 

\begin{align}
	\mathbf{I}_{sdp}=SIM(\mathbf{I}_{dp})
\end{align}
where $SIM(\cdot)$ denotes the simplification process. This simplification is necessary because the boundary between the upper and lower body in the original DensePose map is not robust, causing flickering artifacts. This simplification improves temporal consistency and reduces artifacts, as demonstrated in~\autoref{sec_ablation}.

During inference, we extract the hybrid person representation $\mathbf{I}_{hybrid}$ and feed it to the per-garment network to generate the synthesized target garment image and garment mask. Then, we composite the synthesized target garment image onto the input frame using the predicted garment mask. With the simplified DensePose map, the synthesized garment image aligns accurately with the human body. Therefore, our method does not require any post-processing to achieve accurate human-garment alignment.

\subsection{Per-garment dataset collection with a human body}
\label{sec_dataset}

We employ a Garment Synthesis (GS) network to synthesize target garment images. The input is
 hybrid person representation $\mathbf{I}_{hybrid}$ and the outputs are a target garment image $\mathbf{I}_{g}\in \mathbb{R}^{3\times H \times W}$ and a mask $\mathbf{I}_{m}\in \mathbb{R}^{1\times H \times W}$:

\begin{align}
	\mathbf{I}_{g}, \mathbf{I}_{m} &=GS(\mathbf{I}_{hybrid}) \\
	&=GS(\mathbf{I}_{vm}\oplus SIM(\mathbf{I}_{dp}))
\end{align}

Therefore, to train the GS network, we need to collect a dataset that includes target garment image $\mathbf{I}_{g}$, target garment mask $\mathbf{I}_{m}$, virtual measurement garment image $\mathbf{I}_{vm}$, and the DensePose map $\mathbf{I}_{dp}$.

In contrast to previous research~\cite{chong2021per, wu2024virtual} that utilized a robotic mannequin and took about two hours to collect a per-garment dataset, our method employs real humans for the following reasons: 
\begin{itemize} 
    \item Try-on movements exhibit limited diversity, allowing them to be effectively represented by a small set of predefined poses, which can be performed within a short, tolerable duration for real humans. 
    \item Developing a robotic mannequin that can faithfully replicate human body deformation caused by the interactions between bone, flesh, and skin is challenging and costly. 
    \item Significant differences in appearance between robotic mannequins and real humans make it difficult to use existing human-centric vision models for generating alignment annotations. By using real humans, we can utilize the simplified DensePose map $\mathbf{I}_{sdp}$ as the annotation for alignment.
\end{itemize}

 \noindent\textbf{Data capturing environment setup.}
As shown in~\autoref{fig_teaser}(a), we positioned a camera at chest height and at a sufficient distance to ensure the entire body is captured within the frame. Additionally, we placed a large monitor diagonally in front of the person to display the predefined poses, making it easier for them to follow along without needing to remember all the poses. Natural indoor lighting was used for the video recording. The recording of the participant performing the predefined movements was conducted in 4K resolution at 30 fps, and the entire process took approximately two minutes. 
 
\noindent \textbf{Data capturing procedure.}
We designed a set of predefined poses for the per-garment dataset collection, as illustrated in \autoref{fig_allposes}. There are 14 predefined symmetrical poses in total. Despite the limited number and the absence of asymmetrical poses, experimental results demonstrate that our trained model generalizes well to a wide range of try-on movements.
To collect a dataset for a target garment, we instruct a person to wear the garment and mimic each predefined pose in order. If the garment is occluded by hair, we ask the person to tie his/her hair up.
For each pose, the person slowly turns $360^{\circ}$ in place, allowing us to capture the garment’s appearance from multiple perspectives (\autoref{fig_movements}). Notably, the users only need to perform each pose roughly because we will utilize 3D human pose estimation to obtain the actual pose during dataset generation.

After capturing the video, we generate a per-garment dataset for the target garment by processing the video frame-by-frame, as shown in~\autoref{fig_dataset}. 
First, we employ BEV~\cite{sun2022putting} to estimate a 3D human pose represented by an SMPL mesh and then transform it into an image of a virtual measurement garment.
Then, we apply DensePose~\cite{guler2018densepose} to obtain the DensePose map for each frame. 
Finally, we obtain the target garment image and mask using Graphonomy~\cite{gong2019graphonomy}, an off-the-shelf human parsing tool.

\subsection{Training}
\label{sec_training}

We consider the garment synthesis from the hybrid person representations as an image-to-image translation task. 
We employ the GAN-based image-to-image translation method, pix2pixHD~\cite{wang2018high}, due to its efficiency, which aligns with our requirement for real-time virtual try-on. 
The training pipeline is illustrated in~\autoref{fig_training_inference}(a), where the GS network and discriminator architectures are derived from pix2pixHD~\cite{wang2018high}.

Since we only focus on upper-body garment virtual try-on, utilizing the entire original image for training is inefficient. Therefore, we perform ROI (Region of Interest) extraction to crop the upper-body region into a square image of size $s$. This extraction is straightforward due to the availability of body part segmentation from DensePose maps.
We perform ROI extraction on each image in the per-garment dataset:
\begin{align}
	\begin{cases}
		\mathbf{I}^{roi}_{vm}=ROI_s(\mathbf{I}_{vm}) \\
		\mathbf{I}^{roi}_{sdp}=ROI_s(\mathbf{I}_{sdp}) \\
		\mathbf{I}^{roi}_{g}=ROI_s(\mathbf{I}_{g}) \\
		\mathbf{I}^{roi}_{m}=ROI_s(\mathbf{I}_{m})
	\end{cases}
\end{align}
where $ROI_s(\cdot)$ denotes the image transformation that crops the original image into a square region of interest with size $s$. 
During training, we apply random affine transformations following the $ROI_s(\cdot)$ transformation as a method of data augmentation.

Consequently, the input and output pair data $(\mathbf{I}^{roi}_{hybrid}, \mathbf{I}^{roi}_{gm})$ for training are defined as:
 \begin{align}
	\begin{cases}
		\mathbf{I}^{roi}_{hybrid}=\mathbf{I}^{roi}_{vm}\oplus \mathbf{I}^{roi}_{sdp} \\
		\mathbf{I}^{roi}_{gm}=\mathbf{I}^{roi}_{g}\oplus \mathbf{I}^{roi}_{m}
	\end{cases}
\end{align}

For simplicity, we denote $x=\mathbf{I}^{roi}_{hybrid}$ and $y=\mathbf{I}^{roi}_{gm}$. The object function of the GAN can be expressed as:

\begin{align}
	\mathcal{L}_{GAN}=\mathbb{E}[\log D(x,y)] + \mathbb{E}[\log (1-D(x,GS(x)))]
\end{align}
where $GS(\cdot)$ denotes the Garment Synthesis network, $D(\cdot)$ denotes the discriminator.

Following pix2pixHD~\cite{wang2018high}, we incorporate feature matching loss to stabilize the training process and VGG loss~\cite{ledig2017photo} to enhance the quality of the synthesized images. The feature matching loss is defined as:
\begin{align}
	\mathcal{L}_{FM}=FM(GS(x),y)
\end{align}
where $FM(\cdot)$ denotes the feature matching loss function.
 
The VGG loss can then be defined as:
\begin{align}
\mathcal{L}_{VGG}=VGG(\tilde{\mathbf{I}}^{roi}_{g} \odot \tilde{\mathbf{I}}^{roi}_{m}, \mathbf{I}^{roi}_{g})
\end{align}
where $\tilde{\mathbf{I}}^{roi}_{g}$ represents the synthesized garment image, $\tilde{\mathbf{I}}^{roi}_{m}$ denotes the synthesized garment mask, and $VGG(\cdot)$ denotes VGG loss function.

The full objective function, which combines the GAN loss, feature matching loss, and VGG loss, can be expressed as:
\begin{equation}
	\mathop{\arg}\mathop{\min}\limits_{GS}\left( \left(\mathop{\max}\limits_{D}\mathcal{L}_{GAN}\right)+\lambda_0\mathcal{L}_{FM}+\lambda_1\mathcal{L}_{VGG}\right)
\end{equation}
where $\lambda_0$ and $\lambda_1$ are coefficients that control the importance of the respective loss functions.

\subsection{Inference}
\label{sec_inference}
\autoref{fig_training_inference}(b) illustrates the inference pipeline. 
Given an input image $\mathbf{I}$, we generate the corresponding virtual measurement garment image $\mathbf{I}_{vm}$ and simplified DensePose map $\mathbf{I}_{sdp}$.
We then extract the upper-body regions of $(\mathbf{I}_{vm}, \mathbf{I}_{sdp})$ using the body parts segmentation from the DensePose map and obtain the cropped images $(\mathbf{I}^{roi}_{vm}, \mathbf{I}^{roi}_{sdp})$:
\begin{align}
	\begin{cases}
		\mathbf{I}^{roi}_{vm}=ROI_s(\mathbf{I}_{vm}) \\
		\mathbf{I}^{roi}_{sdp}=ROI_s(\mathbf{I}_{sdp})
	\end{cases}
\end{align}

Next, we concatenate $\mathbf{I}^{roi}_{vm}$ and $\mathbf{I}^{roi}_{sdp}$ to obtain the hybrid person representation $\mathbf{I}^{roi}_{hybrid}$. This hybrid person representation is then fed into the GS network to synthesize the target garment image and garment mask:
\begin{align}
	\tilde{\mathbf{I}}^{roi}_{g}, \tilde{\mathbf{I}}^{roi}_{m}=GS(\mathbf{I}^{roi}_{hybrid})
\end{align}

We then transform the synthesized target garment image and mask into the size of the input image using the inverse transform of the aforementioned ROI extraction:
\begin{align}
	\begin{cases}
		\tilde{\mathbf{I}}_{g}=ROI^{-1}_s(\tilde{\mathbf{I}}^{roi}_{g}) \\
		\tilde{\mathbf{I}}_{m}=ROI^{-1}_s(\tilde{\mathbf{I}}^{roi}_{m})
	\end{cases}
\end{align}

Finally, we compose the synthesized target garment image with the input image using the synthesized target garment mask to obtain the try-on result $\mathbf{I}_o$:
\begin{align}
	\mathbf{I}_o=\mathbf{I}\odot (\mathbf{1}-\tilde{\mathbf{I}}_m)+\tilde{\mathbf{I}}_g\odot \tilde{\mathbf{I}}_m
\end{align}
where $\odot$ denotes element-wise multiplication.
Unlike~\cite{wu2024virtual}, which relies solely on 3D pose estimation for coarse human-garment alignment, our approach integrates DensePose to achieve pixel-level alignment, thereby eliminating the need for post-processing.

\subsection{Human-garment interaction}
As shown in~\autoref{fig_interaction}(a), a previous method by~\citet{chong2021per} requires the user to wear a customized measurement garment to facilitate human-garment interaction. In contrast, our method enables human-garment interaction without the need for such measurement garments, as demonstrated in ~\autoref{fig_interaction}(b). The user can wear any arbitrary garment and manipulate it to interact with the synthesized target garment.

Our method supports this interaction by utilizing the simplified DensePose map, which captures general garment deformation during human-garment interaction, as shown in \autoref{fig_interaction}(b). Although DensePose is designed for dense human pose estimation and aims to be garment-invariant, it does not achieve complete garment invariance. Nevertheless, by leveraging this characteristic, our method enables human-garment interaction without the need for any wearable devices.

\begin{figure*}[t!]
    \centering
    \includegraphics[width=\linewidth]{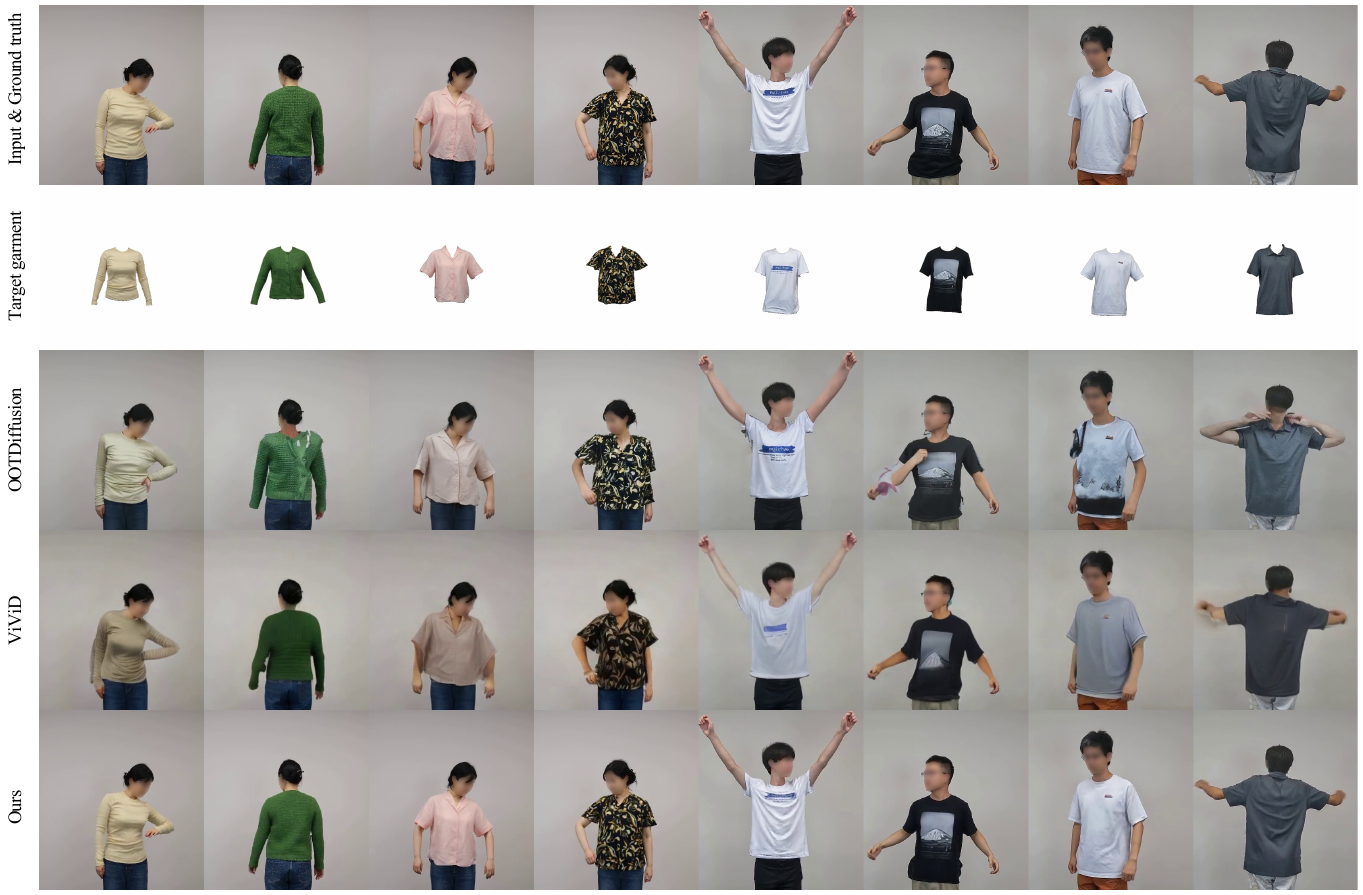}
    \caption{
    \textbf{Qualitative comparison with OOTDiffusion~\cite{xu2024ootdiffusion} and ViViD~\cite{fang2024vivid}.}
    Our method effectively preserves the details of the target garment and produces more realistic try-on results compared to the other two methods. 
    Note that we use the ground-truth image (captured from free-style movements and not included in the training data) shown in the first row as input to demonstrate our method can faithfully recover the real-world wearing result.
    }
    \label{fig_qualitative}
\end{figure*}

\section{Technical Evaluation}
\label{sec_exp}

\subsection{Experiment setting}

\noindent \textbf{Per-garment dataset.} 
We collected $25$ per-garment datasets, including both long-sleeved and short-sleeved garments for males and females. Each dataset contains approximately 3,000 images.
In addition to capturing predefined movements, we recorded a video of the same person wearing the same garment while performing free-style try-on movements lasting about 20 seconds (approximately 600 frames in total). These recordings are intended for both qualitative and quantitative evaluation purposes.

\noindent \textbf{Implementation details.} 
We set the resolution of the region of interest to $512\times512$ and set both $\lambda_0 $ and $\lambda_1$ to $1$.
For each garment, we trained a per-garment model using the corresponding per-garment dataset.
We trained each model on an NVIDIA RTX A6000 GPU for $100$ epochs, with a batch size of $8$ and a learning rate of $2\times10^{-4}$. 
We use the Adam optimizer with $\beta_1=0.5$ and $\beta_2=0.999$. 
The training duration for a garment-specific model was approximately $15$ hours.

\noindent \textbf{Comparison setting.} 
We conduct both qualitative and quantitative comparisons of our method against OOTDiffusion~\cite{xu2024ootdiffusion} and ViViD~\cite{fang2024vivid}, which represent the state-of-the-art in image-based and video virtual try-on techniques. Additionally, we perform a qualitative comparison with \cite{wu2024virtual} regarding human-garment alignment, utilizing their publicly available results due to the non-reproducibility of their method.


\begin{figure*}[h!]
    \centering
    \includegraphics[width=\linewidth]{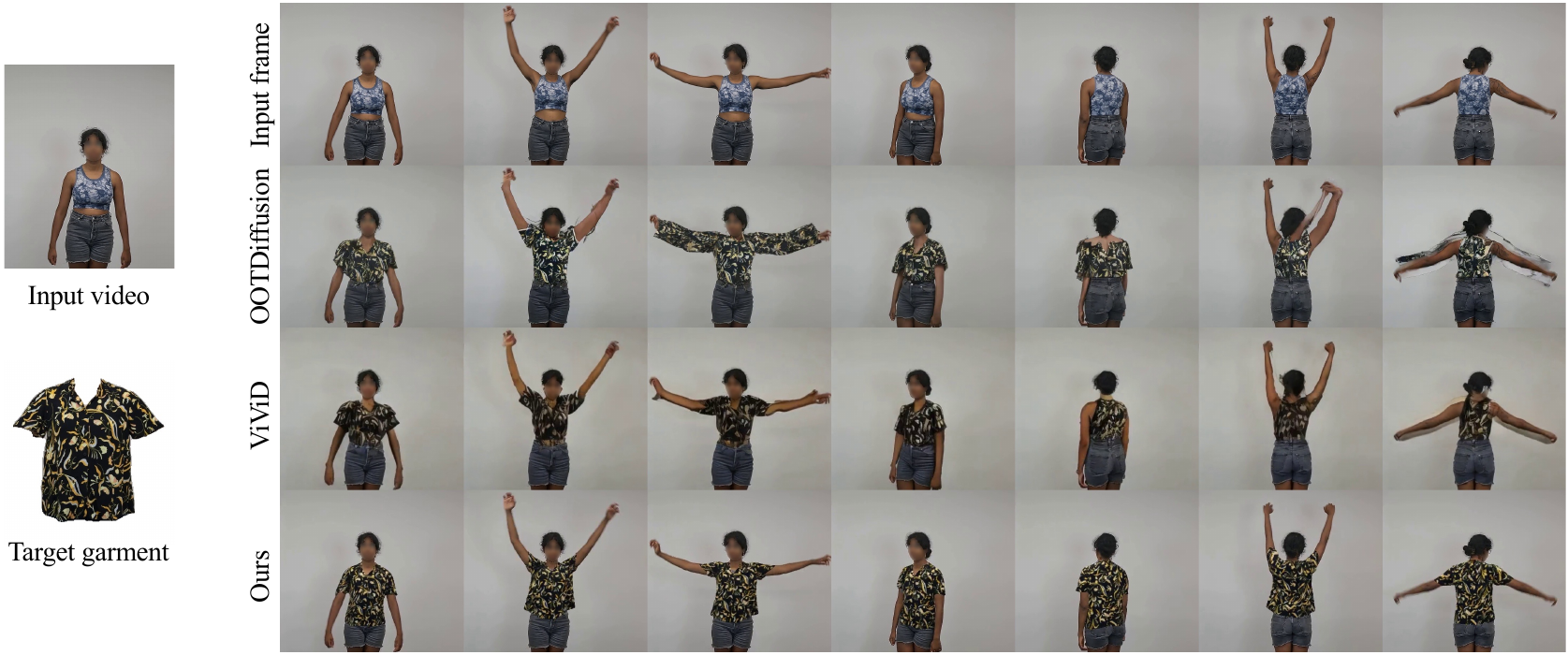}
    \caption{
 \textbf{Temporal consistency comparison with OOTDiffusion~\cite{xu2024ootdiffusion} and ViViD~\cite{fang2024vivid}.}
  Both OOTDiffusion and ViViD generate inconsistent garments, e.g., the target garment becomes long sleeve and no sleeve.
  Moreover, they generate body parts that are not consistent with the input frames.
}
    \label{fig_consistency}
\end{figure*}

\begin{figure}[h!]
    \centering
    \subfloat[\citet{wu2024virtual}]{%
        \includegraphics[height=0.52\linewidth]{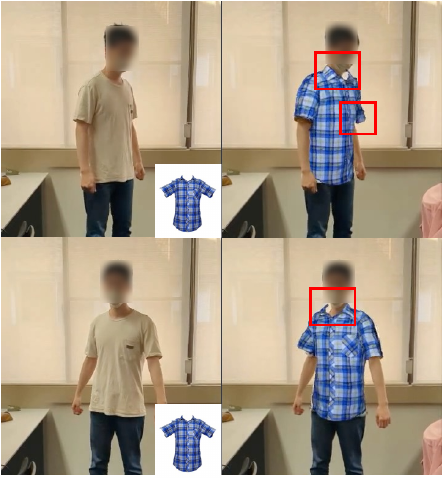}
    }
    \hspace{-0.1cm}
    \subfloat[Ours]{%
        \includegraphics[height=0.52\linewidth]{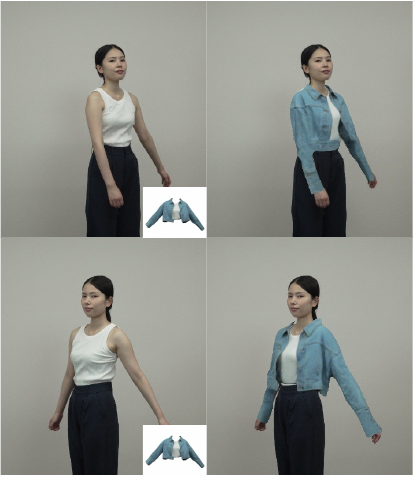}
    }
    \caption{
    Qualitative comparison of human-garment alignment: (a) prior per-garment method (\citet{wu2024virtual}) vs. (b) our method.
    \citet{wu2024virtual} produces noticeable human-garment misalignment, particularly around the collar region.
    }
    \label{fig_alignment_comparison}
\end{figure}
\subsection{Qualitative evaluation}

In~\autoref{fig_qualitative}, we provide a qualitative comparison of image quality among OOTDiffusion~\cite{xu2024ootdiffusion}, ViViD~\cite{fang2024vivid}, and our method. Our method produces more realistic try-on results, whereas the other methods tend to alter the garment characteristics and human identity. Furthermore, we present a qualitative comparison of temporal consistency in~\autoref{fig_consistency}. The results demonstrate that our method consistently generates plausible results across all frames, while the other methods produce varying artifacts in different frames and struggle to maintain temporal consistency. 

Additionally, we compare the human-garment alignment of \citet{wu2024virtual} with our method. Since the results of \cite{wu2024virtual} are non-reproducible, we cannot compare them with the exact same target garment and person. Therefore, we selected a similar garment with collars and similar human body orientations from their supplementary video for comparison. As shown in~\autoref{fig_alignment_comparison}, the results generated by \citet{wu2024virtual} exhibit noticeable human-garment misalignment, particularly around the collar region. In contrast, our method ensures precise human-garment alignment. Please check our supplementary video for more detailed video results.

\subsection{Quantitative evaluation}
To quantitatively evaluate the try-on results generated by our method and previous methods, we leverage the recorded video with free-style try-on movements as both input and ground truth. Although our method supports high-resolution output, for fairness in comparison, we resize the output of all methods to $512 \times 384$, the highest resolution supported by ViViD.
We evaluate the image quality, video quality, and frame rate. 
To measure image quality, we calculated the Structural Similarity Index (SSIM) and Learned Perceptual Image Patch Similarity (LPIPS) between the generated frames and the ground-truth frames.
To measure video quality, we calculated Video Frechet Inception Distance (VFID) using two backbones: I3D~\cite{carreira2017quo} and 3DResNeXt101~\cite{hara2018can}.

As shown in~\autoref{tab_quantitative}(a), our method generates try-on results that outperform all other methods across all metrics.
This result indicates that our method generates high-quality try-on results in real-time and maintains strong temporal consistency.

\setlength{\tabcolsep}{4pt}
\begin{table*}[h!]
\begin{center}
\caption{
\textbf{(a) Quantitative comparison with OOTDiffusion~\cite{xu2024ootdiffusion}, ViViD~\cite{fang2024vivid}, and (b) ablation study on hybrid person representation.}
Metrics marked with $\uparrow$ indicate higher values are better, while those marked with $\downarrow$ indicate lower values are better.
Our method achieves the best performance among all methods on all metrics.
}
\label{tab_quantitative}
\begin{tabular}{c|cccccc}
\toprule[1pt]
\multicolumn{1}{c|}{Experiment} &Method  & SSIM $\uparrow$ & LPIPS $\downarrow$ & $VFID_{I3D}$ $\downarrow$ & $VFID_{ResNeXt}$ $\downarrow$ & fps $\uparrow$\\
\noalign{\smallskip}
\hline
\noalign{\smallskip}
\multirow{3}{*}{(a)} & OOTDiffusion~\cite{xu2024ootdiffusion} & 0.851  & 0.086 &0.459 &0.334 & 0.21\\
 & ViViD ~\cite{fang2024vivid} & 0.831  &0.096 &0.478 &0.365 & 0.89\\
& Ours (\textbf{VM}+\textbf{SDP}) & \textbf{0.920} &\textbf{0.036} &\textbf{0.273} &\textbf{0.184} &\textbf{7.93}\\
\hline
\multirow{3}{*}{(b)}
& \textbf{VM} (Virtual measurement garment only) & 0.902 & 0.043& 0.304&0.216 & -\\
& \textbf{VMDP} (\textbf{VM}+original DensePose map) & 0.905 & 0.043& 0.300&0.208 & -\\
& \textbf{SDP} (Simplified DensePose map only) & 0.904 & 0.044& 0.302&0.207 & -\\
\toprule[1pt]
\end{tabular}
\end{center}
\end{table*}
\setlength{\tabcolsep}{1.4pt}

\begin{figure}[h!]
    \centering
    \includegraphics[width=\linewidth]{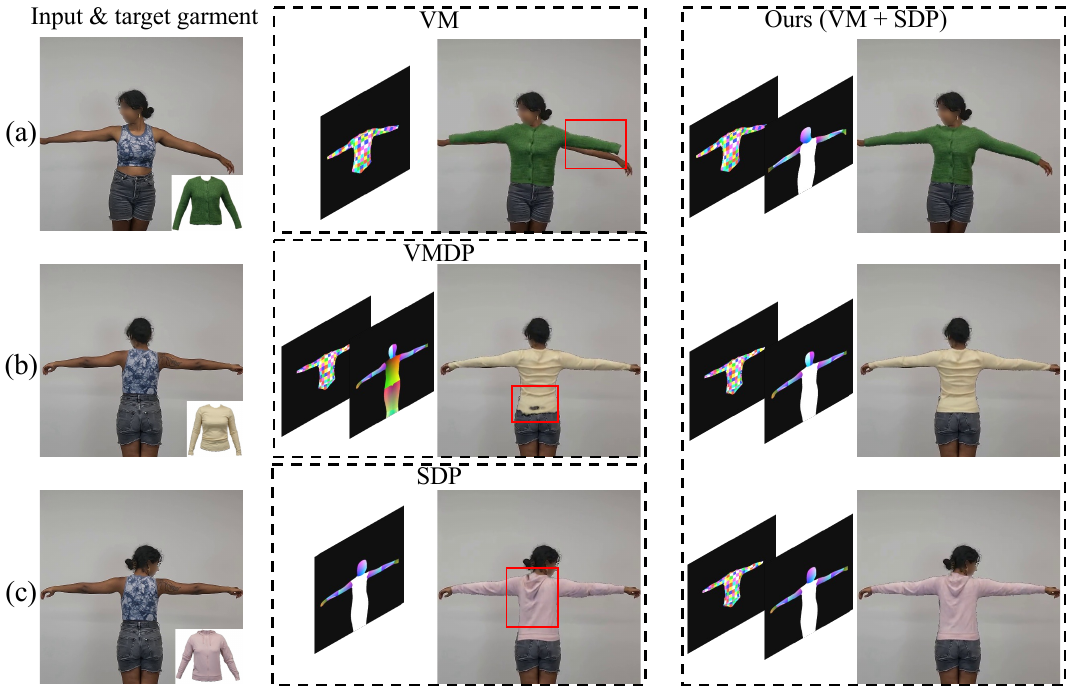}
    \caption{\textbf{Qualitative ablation results for hybrid person representation.} (a) Removing the simplified DensePose map leads to misalignment issues. (b) Using the original DensePose map without simplification introduces artifacts. (c) Omitting the virtual measurement garment results in inconsistent orientation of the synthesized garment with the human body.}
    \label{fig_ablation}
\end{figure}

\begin{figure*}[h!]
    \centering
    \includegraphics[width=\linewidth]{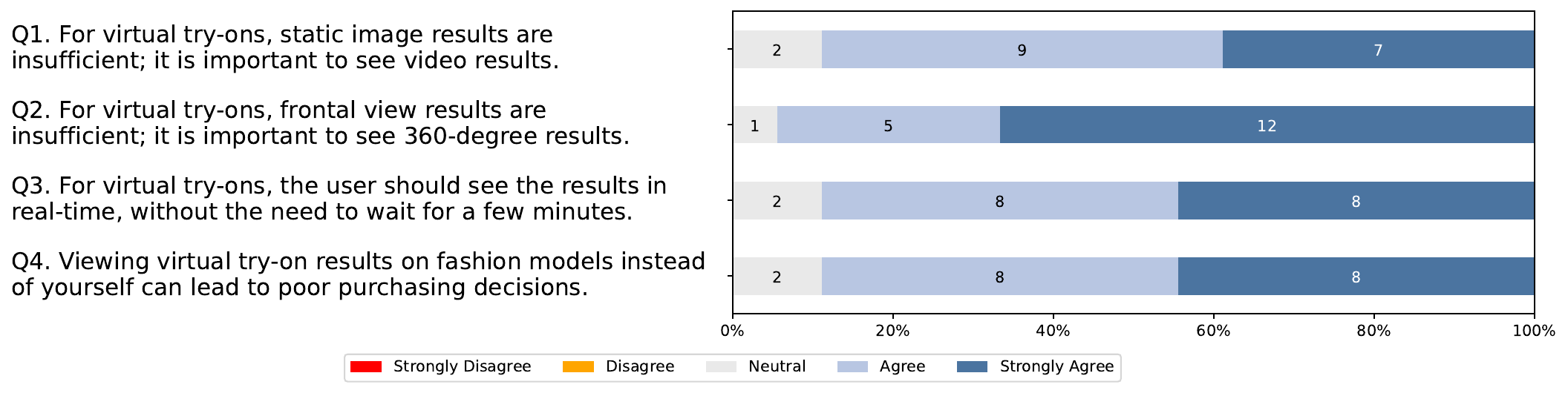}
    \caption{
    The results of the questionnaire asked about the participants' expectations of virtual try-on.
    }
    \label{fig_likert_a}
\end{figure*}

\begin{figure*}[h!]
    \centering
    \includegraphics[width=\linewidth]{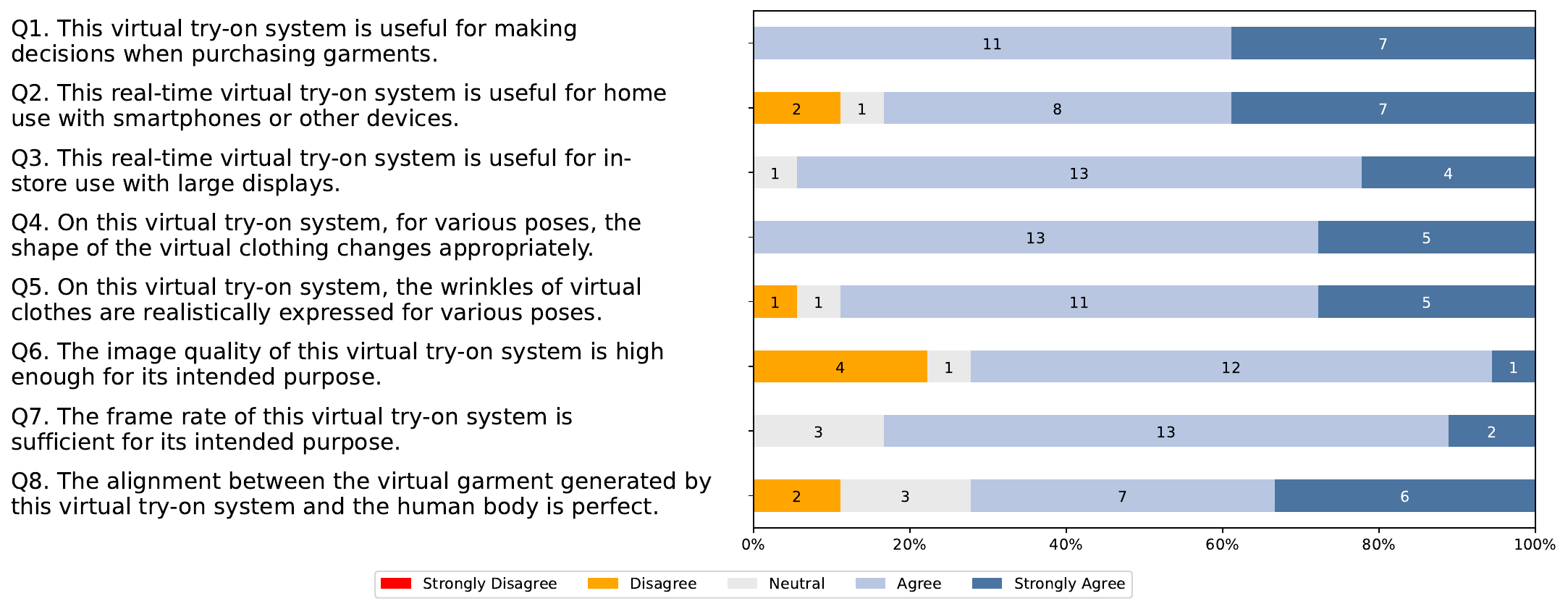}
    \caption{Results of the questionnaire asking the participants' rating on eight statements about the usefulness and user experience of experiencing our virtual try-on system.}
    \label{fig_likert_b}
\end{figure*}

\subsection{Ablation study}
\label{sec_ablation}
To verify the effectiveness of different components in our hybrid person representation, we conduct ablation studies with the following alternative intermediate representations:
\begin{itemize}
	\item \textbf{VM}: Using only the virtual measurement garment as the intermediate representation to demonstrate the benefit of the DensePose map.
	\item \textbf{VMDP}: Using the virtual measurement garment and the original DensePose map as the intermediate representation to show the benefit of simplification of DensePose.
	\item \textbf{SDP}: Using only the simplified DensePose map as the intermediate representation to illustrate the benefit of the virtual measurement garment.

\end{itemize}

\autoref{fig_ablation} illustrates the qualitative ablation results. The simplified DensePose map contributes to precise alignment between the synthesized garment and the human body (\autoref{fig_ablation}(a)). Additionally, simplifying the DensePose map reduces artifacts and enhances the temporal consistency of the synthesized garment by removing non-robust parts of the original DensePose map (\autoref{fig_ablation}(b)). Furthermore, the virtual measurement garment ensures consistent orientation between the synthesized garment and the human body (\autoref{fig_ablation}(c)). As demonstrated in~\autoref{tab_quantitative}(b), our proposed hybrid person representation outperforms other alternative intermediate representations by a large margin.


\section{User Study}
We conducted a user study to evaluate the usability of our dataset collection method and the utility of our interactive virtual try-on system. Participants were asked to try our virtual try-on system and then act as human bodies for dataset collection. We prepared two questionnaires: the first assessed participants' expectations regarding virtual try-on technology, while the second evaluated the performance of our system. We did not include comparisons with OOTDiffusion~\cite{xu2024ootdiffusion} or ViViD~\cite{fang2024vivid}, as these methods are incapable of running in real-time. Additionally, comparisons with previous per-garment methods~\cite{chong2021per,wu2024virtual} were omitted due to the non-reproducibility of their results, which rely on a customized robotic mannequin.

We recruited 18 female participants, aged 18 to 50, all of whom have experience purchasing garments online. These participants were non-experts with no professional training or prior knowledge of the procedure. We specifically chose female participants due to their typically stronger interest in virtual try-on technologies. Additionally, our virtual try-on prototype focuses primarily on female garments, which are generally more diverse and complex.

Each participant spent about 5 minutes trying our system. As shown in~\autoref{fig_teaser}(a), we place a large display and a web camera in front of participants, allowing them to see the virtual try-on results on the display in real-time. They could switch to other virtual garments by touching the screen. After experiencing our virtual try-on system, participants were asked to act as the human body to collect two per-garment datasets. Finally, we asked them to complete the questionnaires. The entire session lasted about 30 minutes, and we compensated the participants for their time.

\subsection{Feedback for dataset collection}
Following the instructions in~\autoref{sec_dataset}, each participant helped collect two per-garment datasets, with each dataset taking about two minutes to collect. Overall, participants completed the dataset collection session without significant difficulty. Only one participant reported feeling ``\textit{a little bit dizzy when rotating in place}''. Therefore, we believe that using real human bodies for dataset collection is feasible, even without prior professional training.

\begin{figure*}[h!]
    \centering
    \subfloat[]{%
        \includegraphics[height=0.21\linewidth]{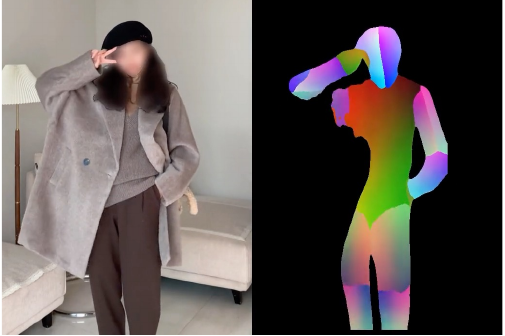}
    }
    \hspace{-0.1cm}
    \subfloat[]{%
        \includegraphics[height=0.21\linewidth]{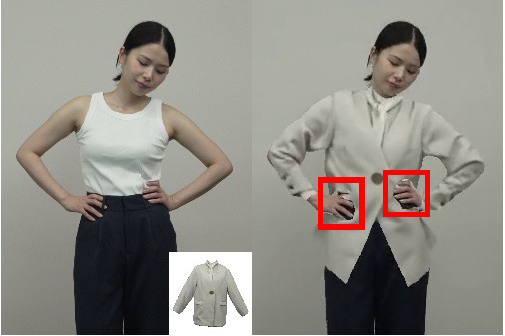}
    }
    \hspace{-0.1cm}
    \subfloat[]{%
        \includegraphics[height=0.21\linewidth]{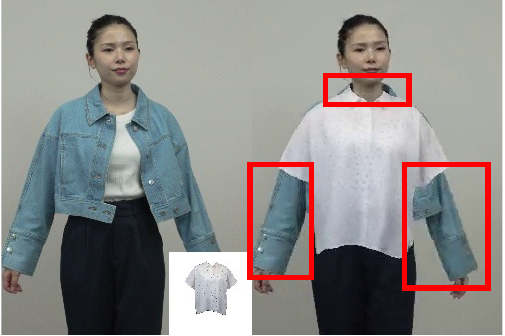}
    }
    \caption{Limitations of our method: (a) Our method is unable to process inputs with extremely loose garments or generate datasets for such garments due to the failure of DensePose map estimation for these inputs; (b) Inaccurate composition mask for hands; (c) Original garment is not removed.}
    \label{fig_failure}
\end{figure*}

\subsection{Feedback for virtual try-on system}

\noindent\textbf{Expectation for virtual try-on.} \autoref{fig_likert_a} presents an overview of the questions and the ratings provided by participants regarding their expectations for virtual try-on. In Q1, "For virtual try-on, static image results are insufficient; it is important to see video results" (Mean=4.28, SD=0.65), most participants agreed on the importance of video results. Additionally, they emphasized the significance of being able to see try-on results in $360^{\circ}$ (Mean=4.61, SD=0.59) and in real-time (Mean=4.33, SD=0.67). In Q4, "Viewing virtual try-on results on a fashion model instead of yourself can lead to poor purchasing decisions" (Mean=4.33, SD=0.67), the majority of participants concurred that using fashion models for virtual try-ons can be misleading. These findings support our approach of prioritizing a customer-oriented virtual try-on by collecting per-garment datasets.

\noindent\textbf{System utility.}
In Q1 of~\autoref{fig_likert_b}, most participants agreed that our system is helpful for making decisions when purchasing garments (Mean=4.39, SD=0.49). Q2 and Q3 reveal that many participants also found our system useful for home use with portable devices (Mean=4.11, SD=0.94) and for in-store use with large displays (Mean=4.17, SD=0.50). 

\noindent\textbf{System performance.} In Q4 and Q5 of~\autoref{fig_likert_b}, most participants agreed that the garment deforms appropriately (Mean=4.28, SD=0.45) and that the wrinkles appear realistic (Mean=4.11, SD=0.74) across various poses. In Q6, while many participants agreed that the image quality is sufficient for virtual try-on (Mean=3.56, SD=0.90), four participants expressed disagreement. We believe this results from our decision to prioritize real-time performance over image quality, as most participants were satisfied with our system's frame rate (Mean=3.94, SD=0.52). Finally, in Q8, most participants agreed that the human-garment alignment is perfect (Mean=3.94, SD=0.97).

Specifically, participant P3 commented, ``\textit{The wrinkles looked realistic. Additionally, the video played smoothly without any lag, providing a comfortable try-on experience.}'' Participant P6 added, ``\textit{This system allows for virtual try-on from various angles, making it practical as you can check the appearance from the sides and back.}''

\section{Limitations and Discussion}
\begin{description}[style=unboxed,leftmargin=0cm]
    \item[Inability to handle extremely loose garments.] Our method relies on the DensePose map for human-garment alignment; however, DensePose map estimation often fails when individuals wear extremely loose garments that occlude the body, as illustrated in~\autoref {fig_failure}(a)\footnote{The fashion model image is sourced from \url{https://v.douyin.com/0hnko66OvwQ/}}. Consequently, we cannot collect per-garment datasets for such garments, and users must avoid wearing them when using our system. In the future, we plan to enhance our intermediate representation to robustly estimate human body shape.
    \item[Inaccurate composition mask.] We utilize the predicted garment mask for composition, which generally performs well. However, as illustrated in \autoref{fig_failure}(b), our method struggles to generate an accurate composition mask for the hands when they are in front of the synthesized garment. This limitation arises from the DensePose map's inability to capture the silhouettes of fingers. Exploring additional representations to capture such fine-grained details is a promising direction for future work.
    \item[Presence of original garment in the input images.] Our method overlays the synthesized garment directly onto the input frame without removing the original garment, as shown in~\autoref{fig_failure}(c). In the future, we plan to develop a skin inpainting module to inpaint the gaps created by the removal of the original garment.
    \item[Necessity of data capture for each garment.] Our method involves capturing a dataset for each garment using real humans. This is certainly an overhead. However, since retailers already hire fashion models for product photos, the additional effort required to collect per-garment datasets is minimal compared to the substantial benefits of personalized, interactive, and  $360^{\circ}$ virtual try-on experiences. Our user study supports this claim, showing that users find our system helpful in making purchase decisions.
    \end{description}
\section{Conclusion}

We present a low-barrier dataset collection method for an interactive per-garment virtual try-on system that delivers high-fidelity, temporally smooth results. Our approach leverages real human bodies to collect per-garment datasets, eliminating the need for an expensive customized robotic mannequin and significantly improving the practical applicability of per-garment virtual try-on techniques.
Additionally, we propose a hybrid person representation for garment synthesis that ensures the synthesized garment images are consistently oriented and precisely aligned with the human body without requiring additional wearable devices. This innovation also facilitates human-garment interaction without the need for extra wearable devices, further improving the practical usability of our method.
We demonstrate the superiority of our approach through qualitative and quantitative evaluations against other state-of-the-art virtual try-on methods. We also validate the necessity of all components of the hybrid person representation through ablation studies. Finally, a user study revealed that most participants found our virtual try-on system helpful for making purchasing decisions.


\bibliographystyle{ACM-Reference-Format}
\bibliography{paper.bib}


\begin{thebibliography}{48}


\ifx \showCODEN    \undefined \def \showCODEN     #1{\unskip}     \fi
\ifx \showISBNx    \undefined \def \showISBNx     #1{\unskip}     \fi
\ifx \showISBNxiii \undefined \def \showISBNxiii  #1{\unskip}     \fi
\ifx \showISSN     \undefined \def \showISSN      #1{\unskip}     \fi
\ifx \showLCCN     \undefined \def \showLCCN      #1{\unskip}     \fi
\ifx \shownote     \undefined \def \shownote      #1{#1}          \fi
\ifx \showarticletitle \undefined \def \showarticletitle #1{#1}   \fi
\ifx \showURL      \undefined \def \showURL       {\relax}        \fi
\providecommand\bibfield[2]{#2}
\providecommand\bibinfo[2]{#2}
\providecommand\natexlab[1]{#1}
\providecommand\showeprint[2][]{arXiv:#2}

\bibitem[Carney et~al\mbox{.}(2020)]%
        {carney2020teachable}
\bibfield{author}{\bibinfo{person}{Michelle Carney}, \bibinfo{person}{Barron Webster}, \bibinfo{person}{Irene Alvarado}, \bibinfo{person}{Kyle Phillips}, \bibinfo{person}{Noura Howell}, \bibinfo{person}{Jordan Griffith}, \bibinfo{person}{Jonas Jongejan}, \bibinfo{person}{Amit Pitaru}, {and} \bibinfo{person}{Alexander Chen}.} \bibinfo{year}{2020}\natexlab{}.
\newblock \showarticletitle{Teachable machine: Approachable Web-based tool for exploring machine learning classification}. In \bibinfo{booktitle}{\emph{Extended abstracts of the 2020 CHI conference on human factors in computing systems}}. \bibinfo{pages}{1--8}.
\newblock


\bibitem[Carreira and Zisserman(2017)]%
        {carreira2017quo}
\bibfield{author}{\bibinfo{person}{Joao Carreira} {and} \bibinfo{person}{Andrew Zisserman}.} \bibinfo{year}{2017}\natexlab{}.
\newblock \showarticletitle{Quo vadis, action recognition? a new model and the kinetics dataset}. In \bibinfo{booktitle}{\emph{proceedings of the IEEE Conference on Computer Vision and Pattern Recognition}}. \bibinfo{pages}{6299--6308}.
\newblock


\bibitem[Casado-Elvira et~al\mbox{.}(2022)]%
        {casado2022pergamo}
\bibfield{author}{\bibinfo{person}{Andr{\'e}s Casado-Elvira}, \bibinfo{person}{Marc~Comino Trinidad}, {and} \bibinfo{person}{Dan Casas}.} \bibinfo{year}{2022}\natexlab{}.
\newblock \showarticletitle{Pergamo: Personalized 3d garments from monocular video}. In \bibinfo{booktitle}{\emph{Computer Graphics Forum}}, Vol.~\bibinfo{volume}{41}. Wiley Online Library, \bibinfo{pages}{293--304}.
\newblock


\bibitem[Chen et~al\mbox{.}(2021)]%
        {chen2021fashionmirror}
\bibfield{author}{\bibinfo{person}{Chieh-Yun Chen}, \bibinfo{person}{Ling Lo}, \bibinfo{person}{Pin-Jui Huang}, \bibinfo{person}{Hong-Han Shuai}, {and} \bibinfo{person}{Wen-Huang Cheng}.} \bibinfo{year}{2021}\natexlab{}.
\newblock \showarticletitle{Fashionmirror: Co-attention feature-remapping virtual try-on with sequential template poses}. In \bibinfo{booktitle}{\emph{Proceedings of the IEEE/CVF International Conference on Computer Vision}}. \bibinfo{pages}{13809--13818}.
\newblock


\bibitem[Choi et~al\mbox{.}(2021)]%
        {choi2021viton}
\bibfield{author}{\bibinfo{person}{Seunghwan Choi}, \bibinfo{person}{Sunghyun Park}, \bibinfo{person}{Minsoo Lee}, {and} \bibinfo{person}{Jaegul Choo}.} \bibinfo{year}{2021}\natexlab{}.
\newblock \showarticletitle{Viton-hd: High-resolution virtual try-on via misalignment-aware normalization}. In \bibinfo{booktitle}{\emph{Proceedings of the IEEE/CVF conference on computer vision and pattern recognition}}. \bibinfo{pages}{14131--14140}.
\newblock


\bibitem[Chong et~al\mbox{.}(2021)]%
        {chong2021per}
\bibfield{author}{\bibinfo{person}{Toby Chong}, \bibinfo{person}{I-Chao Shen}, \bibinfo{person}{Nobuyuki Umetani}, {and} \bibinfo{person}{Takeo Igarashi}.} \bibinfo{year}{2021}\natexlab{}.
\newblock \showarticletitle{Per garment capture and synthesis for real-time virtual try-on}. In \bibinfo{booktitle}{\emph{The 34th Annual ACM Symposium on User Interface Software and Technology}}. \bibinfo{pages}{457--469}.
\newblock


\bibitem[Cirio et~al\mbox{.}(2014)]%
        {cirio2014yarn}
\bibfield{author}{\bibinfo{person}{Gabriel Cirio}, \bibinfo{person}{Jorge Lopez-Moreno}, \bibinfo{person}{David Miraut}, {and} \bibinfo{person}{Miguel~A Otaduy}.} \bibinfo{year}{2014}\natexlab{}.
\newblock \showarticletitle{Yarn-level simulation of woven cloth}.
\newblock \bibinfo{journal}{\emph{ACM Transactions on Graphics (TOG)}} \bibinfo{volume}{33}, \bibinfo{number}{6} (\bibinfo{year}{2014}), \bibinfo{pages}{1--11}.
\newblock


\bibitem[Dong et~al\mbox{.}(2019)]%
        {dong2019fw}
\bibfield{author}{\bibinfo{person}{Haoye Dong}, \bibinfo{person}{Xiaodan Liang}, \bibinfo{person}{Xiaohui Shen}, \bibinfo{person}{Bowen Wu}, \bibinfo{person}{Bing-Cheng Chen}, {and} \bibinfo{person}{Jian Yin}.} \bibinfo{year}{2019}\natexlab{}.
\newblock \showarticletitle{Fw-gan: Flow-navigated warping gan for video virtual try-on}. In \bibinfo{booktitle}{\emph{Proceedings of the IEEE/CVF international conference on computer vision}}. \bibinfo{pages}{1161--1170}.
\newblock


\bibitem[Fang et~al\mbox{.}(2024)]%
        {fang2024vivid}
\bibfield{author}{\bibinfo{person}{Zixun Fang}, \bibinfo{person}{Wei Zhai}, \bibinfo{person}{Aimin Su}, \bibinfo{person}{Hongliang Song}, \bibinfo{person}{Kai Zhu}, \bibinfo{person}{Mao Wang}, \bibinfo{person}{Yu Chen}, \bibinfo{person}{Zhiheng Liu}, \bibinfo{person}{Yang Cao}, {and} \bibinfo{person}{Zheng-Jun Zha}.} \bibinfo{year}{2024}\natexlab{}.
\newblock \showarticletitle{ViViD: Video Virtual Try-on using Diffusion Models}.
\newblock \bibinfo{journal}{\emph{arXiv preprint arXiv:2405.11794}} (\bibinfo{year}{2024}).
\newblock


\bibitem[Gong et~al\mbox{.}(2019)]%
        {gong2019graphonomy}
\bibfield{author}{\bibinfo{person}{Ke Gong}, \bibinfo{person}{Yiming Gao}, \bibinfo{person}{Xiaodan Liang}, \bibinfo{person}{Xiaohui Shen}, \bibinfo{person}{Meng Wang}, {and} \bibinfo{person}{Liang Lin}.} \bibinfo{year}{2019}\natexlab{}.
\newblock \showarticletitle{Graphonomy: Universal human parsing via graph transfer learning}. In \bibinfo{booktitle}{\emph{Proceedings of the IEEE/CVF Conference on Computer Vision and Pattern Recognition}}. \bibinfo{pages}{7450--7459}.
\newblock


\bibitem[Goodfellow et~al\mbox{.}(2014)]%
        {goodfellow2014generative}
\bibfield{author}{\bibinfo{person}{Ian Goodfellow}, \bibinfo{person}{Jean Pouget-Abadie}, \bibinfo{person}{Mehdi Mirza}, \bibinfo{person}{Bing Xu}, \bibinfo{person}{David Warde-Farley}, \bibinfo{person}{Sherjil Ozair}, \bibinfo{person}{Aaron Courville}, {and} \bibinfo{person}{Yoshua Bengio}.} \bibinfo{year}{2014}\natexlab{}.
\newblock \showarticletitle{Generative adversarial nets}.
\newblock \bibinfo{journal}{\emph{Advances in neural information processing systems}}  \bibinfo{volume}{27} (\bibinfo{year}{2014}).
\newblock


\bibitem[Grigorev et~al\mbox{.}(2023)]%
        {grigorev2023hood}
\bibfield{author}{\bibinfo{person}{Artur Grigorev}, \bibinfo{person}{Michael~J Black}, {and} \bibinfo{person}{Otmar Hilliges}.} \bibinfo{year}{2023}\natexlab{}.
\newblock \showarticletitle{Hood: Hierarchical graphs for generalized modelling of clothing dynamics}. In \bibinfo{booktitle}{\emph{Proceedings of the IEEE/CVF Conference on Computer Vision and Pattern Recognition}}. \bibinfo{pages}{16965--16974}.
\newblock


\bibitem[Guan et~al\mbox{.}(2012)]%
        {guan2012drape}
\bibfield{author}{\bibinfo{person}{Peng Guan}, \bibinfo{person}{Loretta Reiss}, \bibinfo{person}{David~A Hirshberg}, \bibinfo{person}{Alexander Weiss}, {and} \bibinfo{person}{Michael~J Black}.} \bibinfo{year}{2012}\natexlab{}.
\newblock \showarticletitle{Drape: Dressing any person}.
\newblock \bibinfo{journal}{\emph{ACM Transactions on Graphics (ToG)}} \bibinfo{volume}{31}, \bibinfo{number}{4} (\bibinfo{year}{2012}), \bibinfo{pages}{1--10}.
\newblock


\bibitem[G{\"u}ler et~al\mbox{.}(2018)]%
        {guler2018densepose}
\bibfield{author}{\bibinfo{person}{R{\i}za~Alp G{\"u}ler}, \bibinfo{person}{Natalia Neverova}, {and} \bibinfo{person}{Iasonas Kokkinos}.} \bibinfo{year}{2018}\natexlab{}.
\newblock \showarticletitle{Densepose: Dense human pose estimation in the wild}. In \bibinfo{booktitle}{\emph{Proceedings of the IEEE conference on computer vision and pattern recognition}}. \bibinfo{pages}{7297--7306}.
\newblock


\bibitem[Halimi et~al\mbox{.}(2023)]%
        {halimi2023physgraph}
\bibfield{author}{\bibinfo{person}{Oshri Halimi}, \bibinfo{person}{Egor Larionov}, \bibinfo{person}{Zohar Barzelay}, \bibinfo{person}{Philipp Herholz}, {and} \bibinfo{person}{Tuur Stuyck}.} \bibinfo{year}{2023}\natexlab{}.
\newblock \showarticletitle{Physgraph: Physics-based integration using graph neural networks}.
\newblock \bibinfo{journal}{\emph{arXiv preprint arXiv:2301.11841}} (\bibinfo{year}{2023}).
\newblock


\bibitem[Hara et~al\mbox{.}(2018)]%
        {hara2018can}
\bibfield{author}{\bibinfo{person}{Kensho Hara}, \bibinfo{person}{Hirokatsu Kataoka}, {and} \bibinfo{person}{Yutaka Satoh}.} \bibinfo{year}{2018}\natexlab{}.
\newblock \showarticletitle{Can spatiotemporal 3d cnns retrace the history of 2d cnns and imagenet?}. In \bibinfo{booktitle}{\emph{Proceedings of the IEEE conference on Computer Vision and Pattern Recognition}}. \bibinfo{pages}{6546--6555}.
\newblock


\bibitem[Jetchev and Bergmann(2017)]%
        {jetchev2017conditional}
\bibfield{author}{\bibinfo{person}{Nikolay Jetchev} {and} \bibinfo{person}{Urs Bergmann}.} \bibinfo{year}{2017}\natexlab{}.
\newblock \showarticletitle{The conditional analogy gan: Swapping fashion articles on people images}. In \bibinfo{booktitle}{\emph{Proceedings of the IEEE international conference on computer vision workshops}}. \bibinfo{pages}{2287--2292}.
\newblock


\bibitem[Jiang et~al\mbox{.}(2022)]%
        {jiang2022clothformer}
\bibfield{author}{\bibinfo{person}{Jianbin Jiang}, \bibinfo{person}{Tan Wang}, \bibinfo{person}{He Yan}, {and} \bibinfo{person}{Junhui Liu}.} \bibinfo{year}{2022}\natexlab{}.
\newblock \showarticletitle{Clothformer: Taming video virtual try-on in all module}. In \bibinfo{booktitle}{\emph{Proceedings of the IEEE/CVF Conference on Computer Vision and Pattern Recognition}}. \bibinfo{pages}{10799--10808}.
\newblock


\bibitem[Kaldor et~al\mbox{.}(2008)]%
        {kaldor2008simulating}
\bibfield{author}{\bibinfo{person}{Jonathan~M Kaldor}, \bibinfo{person}{Doug~L James}, {and} \bibinfo{person}{Steve Marschner}.} \bibinfo{year}{2008}\natexlab{}.
\newblock \showarticletitle{Simulating knitted cloth at the yarn level}.
\newblock In \bibinfo{booktitle}{\emph{ACM SIGGRAPH 2008 papers}}. \bibinfo{pages}{1--9}.
\newblock


\bibitem[Karras et~al\mbox{.}(2024)]%
        {karras2024fashionvdmvideodiffusionmodel}
\bibfield{author}{\bibinfo{person}{Johanna Karras}, \bibinfo{person}{Yingwei Li}, \bibinfo{person}{Nan Liu}, \bibinfo{person}{Luyang Zhu}, \bibinfo{person}{Innfarn Yoo}, \bibinfo{person}{Andreas Lugmayr}, \bibinfo{person}{Chris Lee}, {and} \bibinfo{person}{Ira Kemelmacher-Shlizerman}.} \bibinfo{year}{2024}\natexlab{}.
\newblock \showarticletitle{Fashion-VDM: Video Diffusion Model for Virtual Try-On}.
\newblock \bibinfo{journal}{\emph{arXiv preprint arXiv:2411.00225}} (\bibinfo{year}{2024}).
\newblock


\bibitem[Karras et~al\mbox{.}(2019)]%
        {karras2019style}
\bibfield{author}{\bibinfo{person}{Tero Karras}, \bibinfo{person}{Samuli Laine}, {and} \bibinfo{person}{Timo Aila}.} \bibinfo{year}{2019}\natexlab{}.
\newblock \showarticletitle{A style-based generator architecture for generative adversarial networks}. In \bibinfo{booktitle}{\emph{Proceedings of the IEEE/CVF conference on computer vision and pattern recognition}}. \bibinfo{pages}{4401--4410}.
\newblock


\bibitem[Karras et~al\mbox{.}(2020)]%
        {karras2020analyzing}
\bibfield{author}{\bibinfo{person}{Tero Karras}, \bibinfo{person}{Samuli Laine}, \bibinfo{person}{Miika Aittala}, \bibinfo{person}{Janne Hellsten}, \bibinfo{person}{Jaakko Lehtinen}, {and} \bibinfo{person}{Timo Aila}.} \bibinfo{year}{2020}\natexlab{}.
\newblock \showarticletitle{Analyzing and improving the image quality of stylegan}. In \bibinfo{booktitle}{\emph{Proceedings of the IEEE/CVF conference on computer vision and pattern recognition}}. \bibinfo{pages}{8110--8119}.
\newblock


\bibitem[Kim et~al\mbox{.}(2024)]%
        {kim2024stableviton}
\bibfield{author}{\bibinfo{person}{Jeongho Kim}, \bibinfo{person}{Guojung Gu}, \bibinfo{person}{Minho Park}, \bibinfo{person}{Sunghyun Park}, {and} \bibinfo{person}{Jaegul Choo}.} \bibinfo{year}{2024}\natexlab{}.
\newblock \showarticletitle{Stableviton: Learning semantic correspondence with latent diffusion model for virtual try-on}. In \bibinfo{booktitle}{\emph{Proceedings of the IEEE/CVF Conference on Computer Vision and Pattern Recognition}}. \bibinfo{pages}{8176--8185}.
\newblock


\bibitem[Lahner et~al\mbox{.}(2018)]%
        {lahner2018deepwrinkles}
\bibfield{author}{\bibinfo{person}{Zorah Lahner}, \bibinfo{person}{Daniel Cremers}, {and} \bibinfo{person}{Tony Tung}.} \bibinfo{year}{2018}\natexlab{}.
\newblock \showarticletitle{Deepwrinkles: Accurate and realistic clothing modeling}. In \bibinfo{booktitle}{\emph{Proceedings of the European conference on computer vision (ECCV)}}. \bibinfo{pages}{667--684}.
\newblock


\bibitem[Ledig et~al\mbox{.}(2017)]%
        {ledig2017photo}
\bibfield{author}{\bibinfo{person}{Christian Ledig}, \bibinfo{person}{Lucas Theis}, \bibinfo{person}{Ferenc Husz{\'a}r}, \bibinfo{person}{Jose Caballero}, \bibinfo{person}{Andrew Cunningham}, \bibinfo{person}{Alejandro Acosta}, \bibinfo{person}{Andrew Aitken}, \bibinfo{person}{Alykhan Tejani}, \bibinfo{person}{Johannes Totz}, \bibinfo{person}{Zehan Wang}, {et~al\mbox{.}}} \bibinfo{year}{2017}\natexlab{}.
\newblock \showarticletitle{Photo-realistic single image super-resolution using a generative adversarial network}. In \bibinfo{booktitle}{\emph{Proceedings of the IEEE conference on computer vision and pattern recognition}}. \bibinfo{pages}{4681--4690}.
\newblock


\bibitem[Lee et~al\mbox{.}(2022)]%
        {lee2022high}
\bibfield{author}{\bibinfo{person}{Sangyun Lee}, \bibinfo{person}{Gyojung Gu}, \bibinfo{person}{Sunghyun Park}, \bibinfo{person}{Seunghwan Choi}, {and} \bibinfo{person}{Jaegul Choo}.} \bibinfo{year}{2022}\natexlab{}.
\newblock \showarticletitle{High-resolution virtual try-on with misalignment and occlusion-handled conditions}. In \bibinfo{booktitle}{\emph{European Conference on Computer Vision}}. Springer, \bibinfo{pages}{204--219}.
\newblock


\bibitem[Loper et~al\mbox{.}(2015)]%
        {loper2015smpl}
\bibfield{author}{\bibinfo{person}{Matthew Loper}, \bibinfo{person}{Naureen Mahmood}, \bibinfo{person}{Javier Romero}, \bibinfo{person}{Gerard Pons-Moll}, {and} \bibinfo{person}{Michael~J Black}.} \bibinfo{year}{2015}\natexlab{}.
\newblock \showarticletitle{SMPL: A Skinned Multi-Person Linear Model}.
\newblock \bibinfo{journal}{\emph{Acm Transactions on Graphics}} \bibinfo{volume}{34}, \bibinfo{number}{Article 248} (\bibinfo{year}{2015}).
\newblock


\bibitem[Narain et~al\mbox{.}(2012)]%
        {narain2012adaptive}
\bibfield{author}{\bibinfo{person}{Rahul Narain}, \bibinfo{person}{Armin Samii}, {and} \bibinfo{person}{James~F O'brien}.} \bibinfo{year}{2012}\natexlab{}.
\newblock \showarticletitle{Adaptive anisotropic remeshing for cloth simulation}.
\newblock \bibinfo{journal}{\emph{ACM transactions on graphics (TOG)}} \bibinfo{volume}{31}, \bibinfo{number}{6} (\bibinfo{year}{2012}), \bibinfo{pages}{1--10}.
\newblock


\bibitem[Pan et~al\mbox{.}(2022)]%
        {pan2022predicting}
\bibfield{author}{\bibinfo{person}{Xiaoyu Pan}, \bibinfo{person}{Jiaming Mai}, \bibinfo{person}{Xinwei Jiang}, \bibinfo{person}{Dongxue Tang}, \bibinfo{person}{Jingxiang Li}, \bibinfo{person}{Tianjia Shao}, \bibinfo{person}{Kun Zhou}, \bibinfo{person}{Xiaogang Jin}, {and} \bibinfo{person}{Dinesh Manocha}.} \bibinfo{year}{2022}\natexlab{}.
\newblock \showarticletitle{Predicting loose-fitting garment deformations using bone-driven motion networks}. In \bibinfo{booktitle}{\emph{ACM SIGGRAPH 2022 Conference Proceedings}}. \bibinfo{pages}{1--10}.
\newblock


\bibitem[Patel et~al\mbox{.}(2020)]%
        {patel2020tailornet}
\bibfield{author}{\bibinfo{person}{Chaitanya Patel}, \bibinfo{person}{Zhouyingcheng Liao}, {and} \bibinfo{person}{Gerard Pons-Moll}.} \bibinfo{year}{2020}\natexlab{}.
\newblock \showarticletitle{Tailornet: Predicting clothing in 3d as a function of human pose, shape and garment style}. In \bibinfo{booktitle}{\emph{Proceedings of the IEEE/CVF conference on computer vision and pattern recognition}}. \bibinfo{pages}{7365--7375}.
\newblock


\bibitem[Ramos et~al\mbox{.}(2020)]%
        {ramos2020interactive}
\bibfield{author}{\bibinfo{person}{Gonzalo Ramos}, \bibinfo{person}{Christopher Meek}, \bibinfo{person}{Patrice Simard}, \bibinfo{person}{Jina Suh}, {and} \bibinfo{person}{Soroush Ghorashi}.} \bibinfo{year}{2020}\natexlab{}.
\newblock \showarticletitle{Interactive machine teaching: a human-centered approach to building machine-learned models}.
\newblock \bibinfo{journal}{\emph{Human--Computer Interaction}} \bibinfo{volume}{35}, \bibinfo{number}{5-6} (\bibinfo{year}{2020}), \bibinfo{pages}{413--451}.
\newblock


\bibitem[Santesteban et~al\mbox{.}(2019)]%
        {santesteban2019learning}
\bibfield{author}{\bibinfo{person}{Igor Santesteban}, \bibinfo{person}{Miguel~A Otaduy}, {and} \bibinfo{person}{Dan Casas}.} \bibinfo{year}{2019}\natexlab{}.
\newblock \showarticletitle{Learning-based animation of clothing for virtual try-on}. In \bibinfo{booktitle}{\emph{Computer Graphics Forum}}, Vol.~\bibinfo{volume}{38}. Wiley Online Library, \bibinfo{pages}{355--366}.
\newblock


\bibitem[Santesteban et~al\mbox{.}(2022)]%
        {santesteban2022snug}
\bibfield{author}{\bibinfo{person}{Igor Santesteban}, \bibinfo{person}{Miguel~A Otaduy}, {and} \bibinfo{person}{Dan Casas}.} \bibinfo{year}{2022}\natexlab{}.
\newblock \showarticletitle{Snug: Self-supervised neural dynamic garments}. In \bibinfo{booktitle}{\emph{Proceedings of the IEEE/CVF Conference on Computer Vision and Pattern Recognition}}. \bibinfo{pages}{8140--8150}.
\newblock


\bibitem[Santesteban et~al\mbox{.}(2021)]%
        {santesteban2021self}
\bibfield{author}{\bibinfo{person}{Igor Santesteban}, \bibinfo{person}{Nils Thuerey}, \bibinfo{person}{Miguel~A Otaduy}, {and} \bibinfo{person}{Dan Casas}.} \bibinfo{year}{2021}\natexlab{}.
\newblock \showarticletitle{Self-supervised collision handling via generative 3d garment models for virtual try-on}. In \bibinfo{booktitle}{\emph{Proceedings of the IEEE/CVF Conference on Computer Vision and Pattern Recognition}}. \bibinfo{pages}{11763--11773}.
\newblock


\bibitem[Sekine et~al\mbox{.}(2014)]%
        {sekine2014virtual}
\bibfield{author}{\bibinfo{person}{Masahiro Sekine}, \bibinfo{person}{Kaoru Sugita}, \bibinfo{person}{Frank Perbet}, \bibinfo{person}{Bj{\"o}rn Stenger}, {and} \bibinfo{person}{Masashi Nishiyama}.} \bibinfo{year}{2014}\natexlab{}.
\newblock \showarticletitle{Virtual fitting by single-shot body shape estimation}. In \bibinfo{booktitle}{\emph{Int. Conf. on 3D Body Scanning Technologies}}, Vol.~\bibinfo{volume}{406}. Citeseer, \bibinfo{pages}{413}.
\newblock


\bibitem[Selle et~al\mbox{.}(2008)]%
        {selle2008robust}
\bibfield{author}{\bibinfo{person}{Andrew Selle}, \bibinfo{person}{Jonathan Su}, \bibinfo{person}{Geoffrey Irving}, {and} \bibinfo{person}{Ronald Fedkiw}.} \bibinfo{year}{2008}\natexlab{}.
\newblock \showarticletitle{Robust high-resolution cloth using parallelism, history-based collisions, and accurate friction}.
\newblock \bibinfo{journal}{\emph{IEEE transactions on visualization and computer graphics}} \bibinfo{volume}{15}, \bibinfo{number}{2} (\bibinfo{year}{2008}), \bibinfo{pages}{339--350}.
\newblock


\bibitem[Simard et~al\mbox{.}(2017)]%
        {simard2017machine}
\bibfield{author}{\bibinfo{person}{Patrice~Y Simard}, \bibinfo{person}{Saleema Amershi}, \bibinfo{person}{David~M Chickering}, \bibinfo{person}{Alicia~Edelman Pelton}, \bibinfo{person}{Soroush Ghorashi}, \bibinfo{person}{Christopher Meek}, \bibinfo{person}{Gonzalo Ramos}, \bibinfo{person}{Jina Suh}, \bibinfo{person}{Johan Verwey}, \bibinfo{person}{Mo Wang}, {et~al\mbox{.}}} \bibinfo{year}{2017}\natexlab{}.
\newblock \showarticletitle{Machine teaching: A new paradigm for building machine learning systems}.
\newblock \bibinfo{journal}{\emph{arXiv preprint arXiv:1707.06742}} (\bibinfo{year}{2017}).
\newblock


\bibitem[Song et~al\mbox{.}(2023)]%
        {song2023image}
\bibfield{author}{\bibinfo{person}{Dan Song}, \bibinfo{person}{Xuanpu Zhang}, \bibinfo{person}{Juan Zhou}, \bibinfo{person}{Weizhi Nie}, \bibinfo{person}{Ruofeng Tong}, {and} \bibinfo{person}{An-An Liu}.} \bibinfo{year}{2023}\natexlab{}.
\newblock \showarticletitle{Image-Based Virtual Try-On: A Survey}.
\newblock \bibinfo{journal}{\emph{arXiv preprint arXiv:2311.04811}} (\bibinfo{year}{2023}).
\newblock


\bibitem[Sun et~al\mbox{.}(2022)]%
        {sun2022putting}
\bibfield{author}{\bibinfo{person}{Yu Sun}, \bibinfo{person}{Wu Liu}, \bibinfo{person}{Qian Bao}, \bibinfo{person}{Yili Fu}, \bibinfo{person}{Tao Mei}, {and} \bibinfo{person}{Michael~J Black}.} \bibinfo{year}{2022}\natexlab{}.
\newblock \showarticletitle{Putting people in their place: Monocular regression of 3d people in depth}. In \bibinfo{booktitle}{\emph{Proceedings of the IEEE/CVF Conference on Computer Vision and Pattern Recognition}}. \bibinfo{pages}{13243--13252}.
\newblock


\bibitem[Wang et~al\mbox{.}(2024)]%
        {wang2024mv}
\bibfield{author}{\bibinfo{person}{Haoyu Wang}, \bibinfo{person}{Zhilu Zhang}, \bibinfo{person}{Donglin Di}, \bibinfo{person}{Shiliang Zhang}, {and} \bibinfo{person}{Wangmeng Zuo}.} \bibinfo{year}{2024}\natexlab{}.
\newblock \showarticletitle{MV-VTON: Multi-View Virtual Try-On with Diffusion Models}.
\newblock \bibinfo{journal}{\emph{arXiv preprint arXiv:2404.17364}} (\bibinfo{year}{2024}).
\newblock


\bibitem[Wang et~al\mbox{.}(2018)]%
        {wang2018high}
\bibfield{author}{\bibinfo{person}{Ting-Chun Wang}, \bibinfo{person}{Ming-Yu Liu}, \bibinfo{person}{Jun-Yan Zhu}, \bibinfo{person}{Andrew Tao}, \bibinfo{person}{Jan Kautz}, {and} \bibinfo{person}{Bryan Catanzaro}.} \bibinfo{year}{2018}\natexlab{}.
\newblock \showarticletitle{High-resolution image synthesis and semantic manipulation with conditional gans}. In \bibinfo{booktitle}{\emph{Proceedings of the IEEE conference on computer vision and pattern recognition}}. \bibinfo{pages}{8798--8807}.
\newblock


\bibitem[Wu et~al\mbox{.}(2022)]%
        {wu2022survey}
\bibfield{author}{\bibinfo{person}{Xingjiao Wu}, \bibinfo{person}{Luwei Xiao}, \bibinfo{person}{Yixuan Sun}, \bibinfo{person}{Junhang Zhang}, \bibinfo{person}{Tianlong Ma}, {and} \bibinfo{person}{Liang He}.} \bibinfo{year}{2022}\natexlab{}.
\newblock \showarticletitle{A survey of human-in-the-loop for machine learning}.
\newblock \bibinfo{journal}{\emph{Future Generation Computer Systems}}  \bibinfo{volume}{135} (\bibinfo{year}{2022}), \bibinfo{pages}{364--381}.
\newblock


\bibitem[Wu et~al\mbox{.}(2024)]%
        {wu2024virtual}
\bibfield{author}{\bibinfo{person}{Zaiqiang Wu}, \bibinfo{person}{Jingyuan Liu}, \bibinfo{person}{Long Hin~Toby Chong}, \bibinfo{person}{I-Chao Shen}, {and} \bibinfo{person}{Takeo Igarashi}.} \bibinfo{year}{2024}\natexlab{}.
\newblock \showarticletitle{Virtual Measurement Garment for Per-Garment Virtual Try-On}. In \bibinfo{booktitle}{\emph{Graphics Interface 2024}}.
\newblock


\bibitem[Xiang et~al\mbox{.}(2021)]%
        {xiang2021modeling}
\bibfield{author}{\bibinfo{person}{Donglai Xiang}, \bibinfo{person}{Fabian Prada}, \bibinfo{person}{Timur Bagautdinov}, \bibinfo{person}{Weipeng Xu}, \bibinfo{person}{Yuan Dong}, \bibinfo{person}{He Wen}, \bibinfo{person}{Jessica Hodgins}, {and} \bibinfo{person}{Chenglei Wu}.} \bibinfo{year}{2021}\natexlab{}.
\newblock \showarticletitle{Modeling clothing as a separate layer for an animatable human avatar}.
\newblock \bibinfo{journal}{\emph{ACM Transactions on Graphics (TOG)}} \bibinfo{volume}{40}, \bibinfo{number}{6} (\bibinfo{year}{2021}), \bibinfo{pages}{1--15}.
\newblock


\bibitem[Xu et~al\mbox{.}(2024b)]%
        {xu2024ootdiffusion}
\bibfield{author}{\bibinfo{person}{Yuhao Xu}, \bibinfo{person}{Tao Gu}, \bibinfo{person}{Weifeng Chen}, {and} \bibinfo{person}{Chengcai Chen}.} \bibinfo{year}{2024}\natexlab{b}.
\newblock \showarticletitle{Ootdiffusion: Outfitting fusion based latent diffusion for controllable virtual try-on}.
\newblock \bibinfo{journal}{\emph{arXiv preprint arXiv:2403.01779}} (\bibinfo{year}{2024}).
\newblock


\bibitem[Xu et~al\mbox{.}(2024a)]%
        {xu2024tunnel}
\bibfield{author}{\bibinfo{person}{Zhengze Xu}, \bibinfo{person}{Mengting Chen}, \bibinfo{person}{Zhao Wang}, \bibinfo{person}{Linyu Xing}, \bibinfo{person}{Zhonghua Zhai}, \bibinfo{person}{Nong Sang}, \bibinfo{person}{Jinsong Lan}, \bibinfo{person}{Shuai Xiao}, {and} \bibinfo{person}{Changxin Gao}.} \bibinfo{year}{2024}\natexlab{a}.
\newblock \showarticletitle{Tunnel Try-on: Excavating Spatial-temporal Tunnels for High-quality Virtual Try-on in Videos}.
\newblock \bibinfo{journal}{\emph{arXiv preprint arXiv:2404.17571}} (\bibinfo{year}{2024}).
\newblock


\bibitem[Zhou and Yatani(2022)]%
        {zhou2022gesture}
\bibfield{author}{\bibinfo{person}{Zhongyi Zhou} {and} \bibinfo{person}{Koji Yatani}.} \bibinfo{year}{2022}\natexlab{}.
\newblock \showarticletitle{Gesture-aware interactive machine teaching with in-situ object annotations}. In \bibinfo{booktitle}{\emph{Proceedings of the 35th Annual ACM Symposium on User Interface Software and Technology}}. \bibinfo{pages}{1--14}.
\newblock


\bibitem[Zhu et~al\mbox{.}(2017)]%
        {zhu2017unpaired}
\bibfield{author}{\bibinfo{person}{Jun-Yan Zhu}, \bibinfo{person}{Taesung Park}, \bibinfo{person}{Phillip Isola}, {and} \bibinfo{person}{Alexei~A Efros}.} \bibinfo{year}{2017}\natexlab{}.
\newblock \showarticletitle{Unpaired image-to-image translation using cycle-consistent adversarial networks}. In \bibinfo{booktitle}{\emph{Proceedings of the IEEE international conference on computer vision}}. \bibinfo{pages}{2223--2232}.
\newblock


\end{thebibliography}



\end{document}